\documentclass[
  aps,
  prl,
  a4paper,
  english,
  twocolumn,
  superscriptaddress
]{revtex4-1}
\usepackage[T1]{fontenc}
\usepackage[utf8]{inputenc}
\usepackage{graphicx}
\usepackage[american]{babel}
\usepackage{amssymb,amsmath,comment}
\usepackage[
  caption=false
]{subfig}
\usepackage{refstyle}
\usepackage[
  citecolor=blue,
  colorlinks,
  linkcolor=blue,
  urlcolor=blue,
]{hyperref}
\usepackage[normalem]{ulem}
\usepackage{xcolor}
\usepackage[babel=true]{csquotes}

\begin{document}

\global\long\def\pgr{\mathcal{P}_{\text{gr}}}
\global\long\def\pdb{\mathcal{P}_{\text{db}}}
\global\long\def\pov{\mathcal{P}_{\text{ov}}}
\global\long\def\pn{\mathcal{P}_{0}}
\global\long\def\df{d_{\text{f}}}

\title{Controlling percolation with limited resources} 

\author{Malte Schr\"oder}
\email{malte.schroeder@ds.mpg.de}
\affiliation{
  Network Dynamics,
  Max Planck Institute for Dynamics and Self-Organization (MPIDS),
  37077 G\"ottingen, Germany
}

\author{Nuno A. M. Ara\'ujo}
\email{nmaraujo@fc.ul.pt}
\affiliation{
  Departamento de F\'isica, Faculdade de Ci\^encias, Universidade de Lisboa, P-1749-016 Lisboa, Portugal
}
\affiliation{
  Centro de F\'isica Te\'orica e Computacional, Universidade de Lisboa, 1749-016 Lisboa, Portugal
}

\author{Didier Sornette}
\email{dsornette@ethz.ch}
\affiliation{Department of Management, Technology and Economics (D-MTEC), ETH Z\"urich, 
Scheuchzerstrasse 7, CH-8092 Zurich, Switzerland}

\author{Jan Nagler}
\email{jnagler@ethz.ch}
\affiliation{Computational Physics for Engineering Materials, Institute for Building Materials, ETH Z\"urich, 
Wolfgang-Pauli-Strasse 27, HIT, CH-8093 Zurich, Switzerland
}

\begin{abstract}
Connectivity - or the lack thereof - is crucial for the function of many man-made systems, from financial and economic networks over epidemic spreading in social networks to technical infrastructure. Often, connections are deliberately established or removed to induce, maintain, or destroy global connectivity. Thus, there has been a great interest in understanding how to control percolation, the transition to large-scale connectivity. Previous work, however, studied control strategies assuming unlimited resources.
Here, we depart from this unrealistic assumption and consider the effect of limited resources on the effectiveness of control. We show that, even for scarce resources, percolation can be controlled with an efficient intervention strategy. We derive this strategy and study its implications, revealing a discontinuous transition as an unintended side-effect of optimal control.
\end{abstract}

\maketitle

We are living in a globalized world. Large-scale connectivity, in particular, is essential for the proper functioning of many socio-economic and technical systems. Examples include technical networks like the internet \cite{Cohen00_resielience, Cohen01_attack, buldyrev10_interdependent} or the world aviation network \cite{verma16_emergence} and a wide range of socio-economic and financial systems \cite{gai10_financial_contagion, helbing13_networked_risk, elliott14_financial}. In other cases connectivity may be a liability, allowing the spreading of diseases and other contagion processes \cite{moore00_epidemic_network, satorras01_epidemic, cohen03_network_immunization}. Ideally, control of connectivity has the goal to prevent wide-spread failure, for example by immunizing a subset of the population to prevent an epidemic. Identifying efficient strategies that use minimal resources is an ongoing problem \cite{pastor02_immunization, chen08_networkImmunization, morone15_influence}. In many cases, however, one cannot completely prevent an undesirable transition, such as a recession or financial crisis, and tries to delay it as long as possible, often resulting in more severe consequences when the transition inevitably occurs \cite{helbing13_networked_risk, lee17_universal, sornette14_perpetualMoney}. Thus, it is essential to understand how to control and delay the emergence of connectivity under the constraint of limited resources and what such unintended consequences may be.

Percolation theory describes the emergence or breakdown of global connectivity depending on the structure of the underlying network with stochastic link addition processes \cite{stauffer_92_percolation_book, callaway00_percolation_random_graphs, cho09_explosiveScalefree, cohen2010complex, newman_2010_networks}.
A large body of work has studied the impact of an unlimited number of small interventions in modified models of network growth with the goal to delay the percolation transition. Most of these processes are based on a specific link addition rule. Typically, two (or more) possible candidate  links are evaluated at each step and the link is added that delays (or enhances) the percolation transition the most \cite{achlioptas09_explosive}. This ``competitive'' percolation \cite{nagler11_single_links} leads to an extremely sudden, but still continuous transition, sometimes referred to as ``explosive'' \cite{costa10_explosive, riordan11_explosive, nagler11_single_links}. Other models introduce explicit control over the largest cluster, which further delays the transition and can result in a genuine discontinuous percolation transition \cite{araujo10_explosive_largest, schrenk11_gaussian, chen13_superciritcal_explosive_bfw, cho13_avoiding_spanning}. Many more models with similar motivation have been studied, leading to a surprising diversity of phenomena \cite{achlioptas09_explosive, cho09_explosiveScalefree, radicchi09_explosiveScalefree, friedman09_explosivepercolation, costa10_explosive, grassberger11_explosive, nagler11_single_links, riordan11_explosive, riordan12_selfaveraging, schrenk12_bfw_lattice, chen13_unstable_supercritical_bfw, schroder13_crackling_percolation, schroeder16_supercritical, dsouza15_review}.

In all these examples control is inherent to the link addition process, implicitly assuming unlimited resources and allowing indefinite control. 
Control in realistic settings, however, will be restricted by limited resources. Here, we derive an efficient resource-limited control strategy to delay percolation and discuss the consequences for the resulting percolation transition. In particular, while the delayed transition remains smooth for sub-optimal interventions, optimizing the control parameters to maximize the delay results in a discontinuous transition.\\

\section*{Results}

\begin{figure*}[t]
\centering
\includegraphics[width=0.65\textwidth]{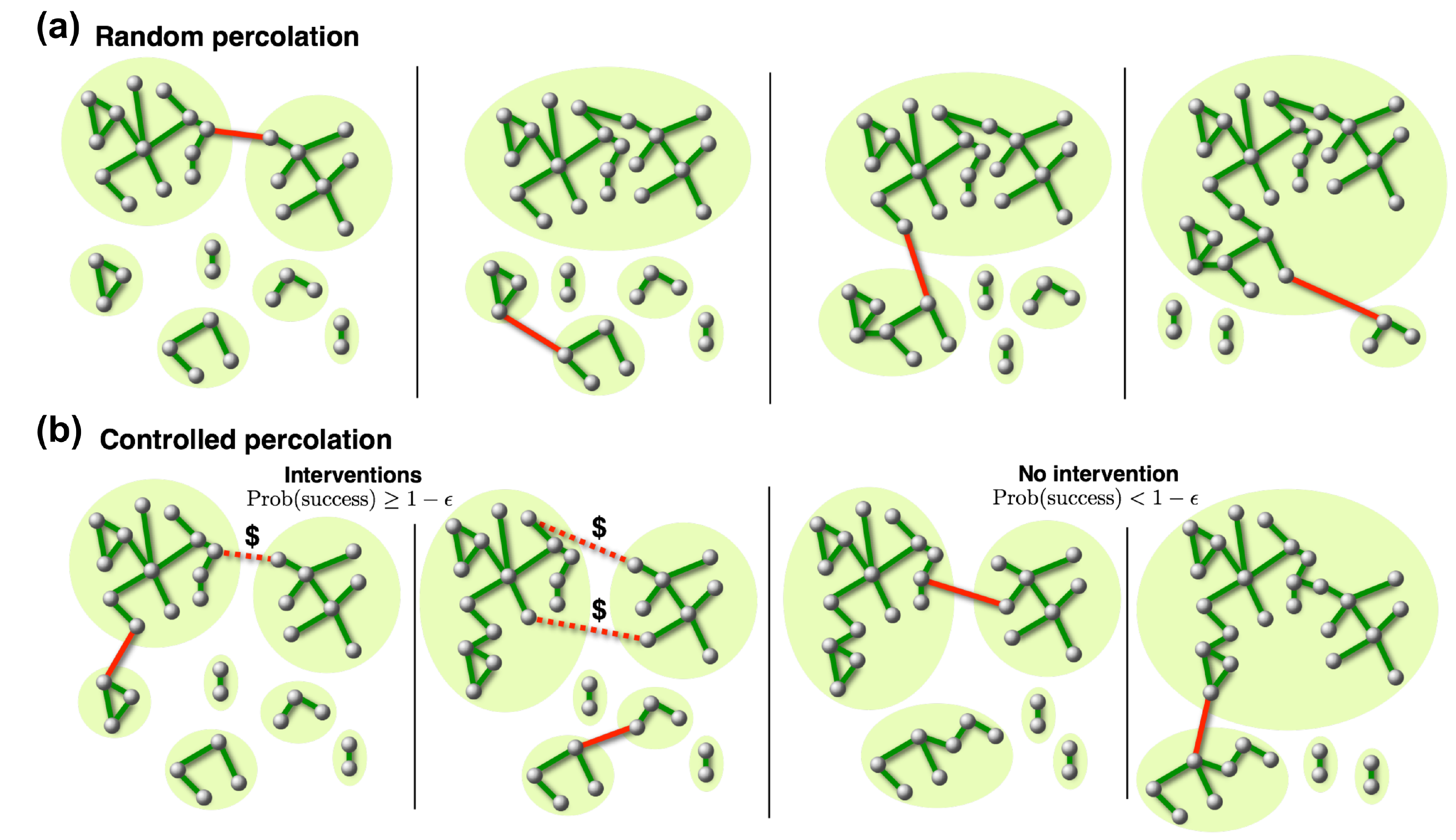}
\caption{\textbf{Controlling the percolation transition.} (a) Random percolation: in each step a link is selected uniformly at random and added to the network. (b) Controlled percolation: In each step, we can prevent the chosen link from being added to the network, paying a cost $c\left[S(i), S(j)\right]$ from a limited budget $B$. The constraint of a limited budget requires an efficient control strategy. As described in the text, we only prevent links when the probability that the intervention is successful is sufficiently large, $\mathrm{Prob}(\mathrm{success}) \ge 1 - \epsilon$. We consider an intervention unsuccessful if a similarly large cluster is likely to appear again with the next link $e_{kl}$. Consequently, we intervene when the probability of such a failure $\mathrm{Prob}(\mathrm{failure}) \approx \mathrm{Prob}\left[S(k) + S(l) \ge S_{ij}\right] < \epsilon$ is small (the expected time until a similarly large clusters appears is large). When this failure probability is too large or the budget is exhausted, we do not intervene. As illustrated, this control delays the creation of large clusters and the onset of percolation.
}
\label{fig:fig1_model_schematic}
\end{figure*}

\noindent\textbf{Model.}
We develop our framework to efficiently delay the percolation transition based on the prototypical model of classical network formation, percolation of a random graph:
new links $e_{ij}$ between nodes $i$ and $j$ are chosen uniformly at random and sequentially added to a set of $N$ initially unconnected nodes \cite{erdos60_evolution_random_graphs}. We implement control of link addition by preventing the chosen link from being added (see Fig.~\ref{fig:fig1_model_schematic}). This control is costly and preventing a link incurs a cost $c\left[S(i), S(j)\right]$, where $S(i)$ and $S(j)$ are the sizes of the respective connected components (clusters) that include the nodes $i$ and $j$. Once a total budget $B$ is spent, we can no longer control the link addition process. We track the evolution of the relative size of the largest connected component $S_1/N$ as a function of the link density $p = L/N$, where $L$ is the number of links added to the network. For the results presented here, the cost of an intervention is kept constant $c\left[S(i), S(j)\right] = 1$ and we assume a budget that scales linearly with the number of nodes, $B=bN$, where $b$ is a (finite) constant. Corresponding results are obtained for other cost functions that scale with the size of the clusters, such as $c\left[S(i), S(j)\right] = S(i) + S(j)$ (see Supplementary Information). In this case, avoiding the transition completely would clearly require preventing most of the links, which is impossible with limited resources.

In order to efficiently utilize the available resources and decide which links to prevent, we derive a control protocol based on the effect of a single intervention. Consider preventing a link $e_{ij}$ that, when added to the network, would create a cluster of size $S_{ij} = S(i) + S(j)$. If the next link $e_{kl}$ creates a cluster of size $S_{kl} = S(k) + S(l) \ge S_{ij}$, we spent some of our budget in vain, since we did not delay the emergence of a large cluster. Conversely, we can consider the intervention effective, when the next links $e_{kl}$ only create smaller clusters $S_{kl} < S_{ij}$ and the emergence of a large cluster was delayed.
Based on this idea we propose a control protocol where we prevent a link $e_{ij}$ only if the expected impact is sufficiently large. We measure this impact by the (expected) number of links $\Delta L_{S_{ij}}$ until a cluster of size at least $S_{ij}$ appears again. Clearly, if $\Delta L_{S_{ij}}$ is large, the intervention is more likely to delay the growth of a large cluster. If this delay is larger than some threshold $\Delta L_\mathrm{thres}$, we consider the intervention effective and prevent the link, otherwise we do not intervene. 
In practice, we estimate the expected $\Delta L_{S_{ij}}$ from the current cluster-size distribution $n_S$ as the inverse of the probability that a new link $e_{kl}$ creates a cluster $S_{kl} \ge S_{ij}$, 
\begin{eqnarray}
\frac{1}{\left<\Delta L_{S_{ij}}\right>} & \approx & \mathrm{Prob}\left[S_{kl} = S(k) + S(l) \ge S_{ij}\right] \\
&=& \sum_{\substack{S(l) \neq S(k)\\ S(k) + S(l) \ge S_{ij}}} \frac{S(k) n_{S(k)}}{N} \times \frac{S(l) n_{S(l)}}{ N - 1 } \nonumber\\
	 &+& \;\; \sum_{2S(k) \ge S_{ij}} \frac{S(k) n_{S(k)}}{N} \times \frac{S(k) (n_{S(k)} - 1)}{ N - 1 }\,, \nonumber
\end{eqnarray}
where the first sum describes the probability of a merger of clusters of different size resulting in a cluster at least as large as $S_{ij}$ and the second sum describes similar mergers between clusters with equal size. For simplicity, we ignore that a link already present cannot be added again.
Hence, we prevent a link from being added if $\mathrm{Prob}\left[S(k) + S(l) \ge S_{ij}\right] < 1/\Delta L_\mathrm{thres} := \epsilon$, where $\epsilon$ denotes the ``intervention intensity'', which is the expected link rejection rate. This protocol is equivalent to stopping the $\epsilon$-fraction most extreme events during the percolation process given sufficient budget. Other control strategies based for example on constraining the variance of the cluster size distribution are less efficient but give qualitatively similar results (see Supplementary Information).\\

\begin{figure}
\centering
\includegraphics[width=0.4\textwidth]{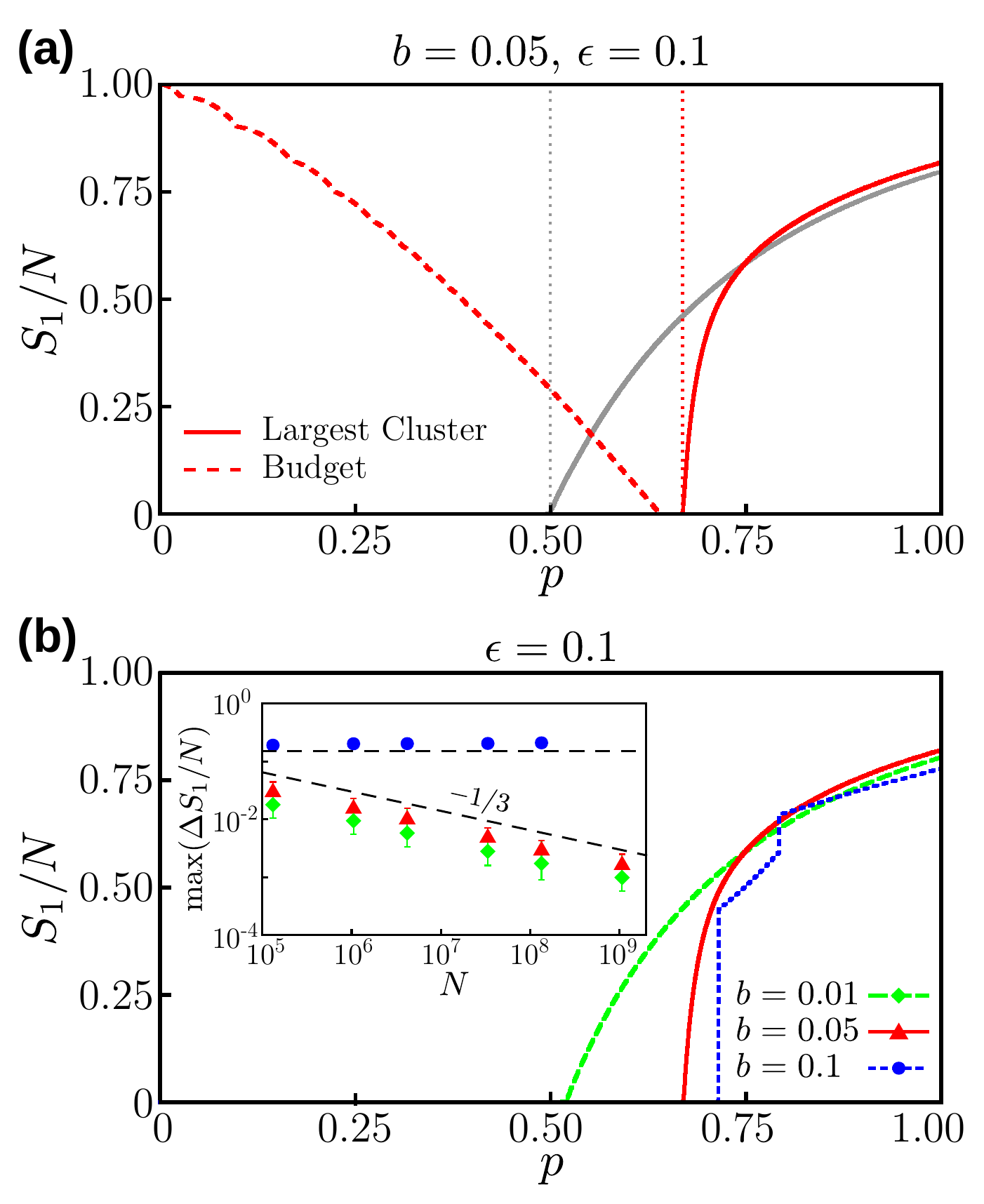}
\caption{\textbf{Effects of resource limited control of percolation.}
(a) Single realization of the evolution of the relative size of the largest cluster for $N=2^{25}$ (red solid line) and remaining fraction of the budget (red dashed line) for budget parameter $b=0.05$ and intervention intensity $\epsilon = 0.1$. Compared to zero budget, the percolation threshold is shifted from $p_c=0.5$ (gray line, showing random percolation without control) to $p_c \approx 0.67$. Interestingly, the transition remains continuous and in the same universality class.
(b) Single realizations of the evolution of the relative size of the largest cluster for $N=2^{25}$, $\epsilon = 0.1$ and different values of $b$. Surprisingly, when $b$ becomes large enough, the transition becomes discontinuous. 
Inset: the largest gap $\max(\Delta S_1/N)$, averaged over $2^{10}$ to $2^6$ realizations. For small $b$, the scaling is the same as expected for random percolation, $\max(\Delta S_1/N) \sim N^{-1/3}$. However, for a sufficiently large budget, the largest gap is independent of the network size and the transition is discontinuous.}
\label{fig:fig2_limited_budget_realizations}
\end{figure}

\noindent\textbf{Efficient control of percolation.}
How much and how efficiently can the percolation transition be delayed with limited resources? As shown in Fig.~\ref{fig:fig2_limited_budget_realizations}, even with a small budget $B=bN=0.05N$, meaning less than one intervention in ten link additions until $p_c \ge 1/2$, we can significantly delay the percolation transition compared to random percolation. Compared to the sudden transitions in the models of explosive percolation \cite{achlioptas09_explosive,costa10_explosive,riordan11_explosive,nagler11_single_links,grassberger11_explosive, dsouza15_review}, our control protocol is more effective in delaying the transition. Interestingly, the transition remains smooth and still belongs to the same universality class as random percolation when the budget is exhausted before the transition (see Supplementary Information for a finite size scaling analysis).

Note that in Fig.~\ref{fig:fig2_limited_budget_realizations}a the budget runs out at \mbox{$p=:p_\mathrm{last}<p_c$}, before the percolation threshold $p_c$, and the transition itself is uncontrolled. We can estimate how long the budget lasts: With a constant intervention rate $\epsilon$ we would expect $\Delta L_\mathrm{int} = \epsilon \Delta L_\mathrm{total}$ interventions to occur during the sampling of $\Delta L_\mathrm{total}$ links. During this period, we add only $N \Delta p = \Delta L = (1-\epsilon) \Delta L_\mathrm{total}$ links. Taking $\Delta L_\mathrm{int} = \Delta B = N \Delta b$, we find the budget used in this interval $\Delta b = \frac{\epsilon}{1-\epsilon}\Delta p$.

However, the budget decays nonlinearly, as seen in Fig.~\ref{fig:fig2_limited_budget_realizations}a, which means the true intervention rate also varies with $p$.
This nonlinear dependency results from the behaviour of the intervention rate oscillating around an effective linear increase $\epsilon_\mathrm{eff}(p) = \min\left[\epsilon, \epsilon\left(1+p/p_c^\mathrm{max}\right)/2\right]$, where $p_c^\mathrm{max}$ is the position of the critical point of controlled percolation with intervention intensity $\epsilon$ and infinite budget (see Supplementary Information for details).
This observation, together with integration over $p$, then yields the closed expression defining $p_\mathrm{last}$
\begin{equation}\label{plast}
b = \int_0^{p_\mathrm{last}} \frac{\epsilon_\mathrm{eff}(p)}{1 - \epsilon_\mathrm{eff}(p)} \mathrm{d}p \,.
\end{equation}
As expected, a larger (effective) intervention rate requires a larger budget. Consequently, for a small budget, (i) the budget runs out before the onset of percolation at $p_\mathrm{last} < p_c$ (interventions stop), (ii) the process is uncontrolled in a short but extensive window prior to the transition point, and (iii) one observes a continuous transition in the same universality class as random percolation. In contrast to previous percolation rules where delaying the transition changes its universality class, the limited resources in our model are exhausted before the transition. At this point the largest cluster has a fixed finite size and uncontrolled random percolation takes over, resulting in a continuous transition similar to random percolation for different initial cluster-size distributions \cite{cho10_percolation_initialcondition}. \\

\begin{figure}
\centering
\includegraphics[width=0.4\textwidth]{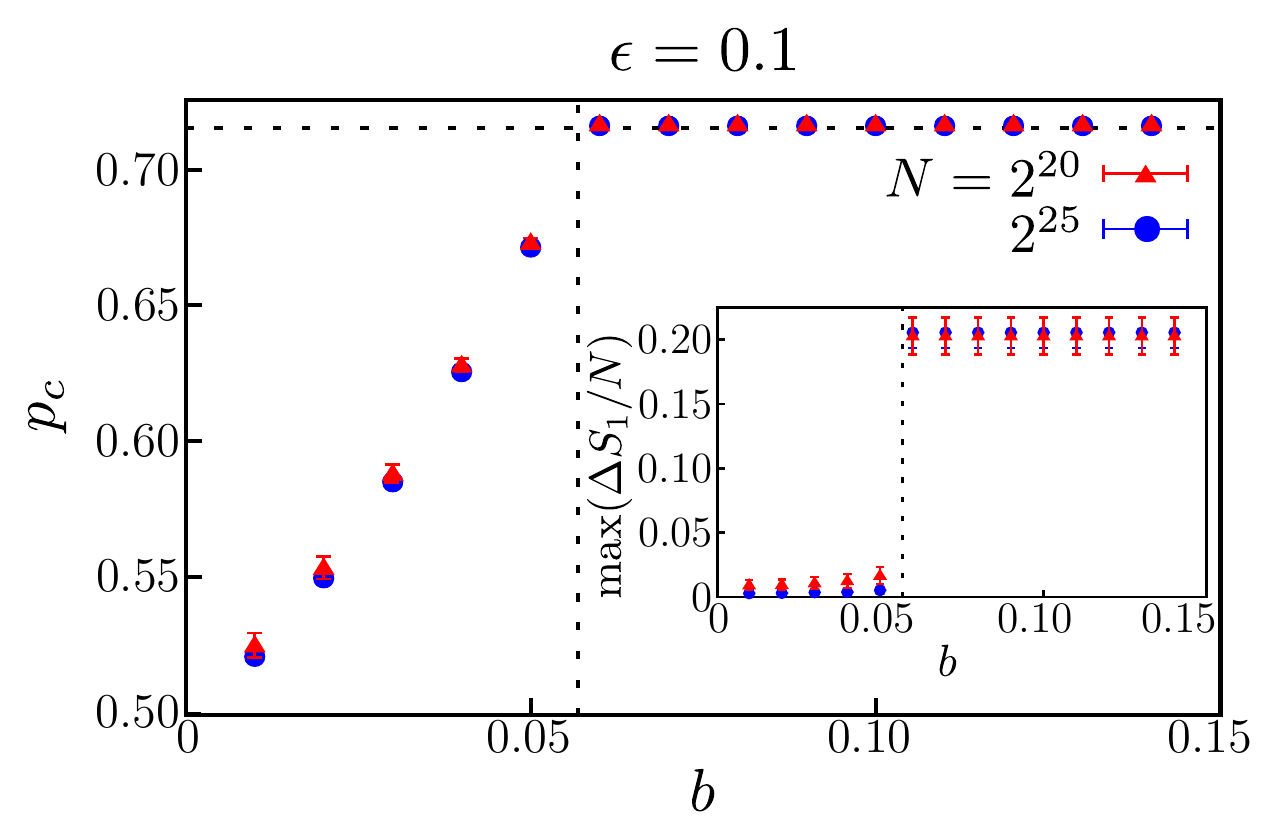}
\caption{\textbf{Discontinuous transition above a critical budget.}
Percolation threshold $p_c$ measured by the position of the largest gap of $S_1$ for different values of $b$. Results are averaged over $1024$ and $256$ realizations for networks of size $N = 2^{20}$ and $2^{25}$, respectively. Error bars indicate the standard deviation. The delay increases with an increasing budget until it becomes constant above a critical budget $b_c \approx 0.058$. At the same time, the transition changes from continuous to discontinuous at $b = b_c$. Inset: The size of the largest gap $\mathrm{max}(\Delta S_1/N)$ for different $b$.
}
\label{fig:fig3_critical_budget_parameters}
\end{figure}

\noindent\textbf{Optimal control leads to discontinuity.}
Increasing the budget also increases the delay of the transition. Interestingly, too large a budget also leads to a discontinuous transition (see Fig.~\ref{fig:fig2_limited_budget_realizations}b). At the same time, increasing the budget further does no longer increase the delay of the transition and $p_c$ becomes constant. 
Clearly, when the budget survives the percolation threshold, additional interventions have no effect on the transition. This suggests that the optimal delay is achieved for an optimal budget lasting exactly until the percolation threshold, $p_\mathrm{last} = p_c$. At the same time, no uncontrolled window exists before the transition and the transition becomes discontinuous.

A similar logic defines the optimal parameters for speeding up the percolation transition (see Supplementary Information): 
interventions taken after the transition have no effect while intervention-free uncontrolled link addition will reduce the effect of previous interventions. Optimal interventions necessarily end exactly at the percolation threshold, regardless of the intended result of the control. 

Substituting $p_\mathrm{last} = p_c^\mathrm{max} \approx 0.72$ in Eq.(\ref{plast}) as the largest observed value of the critical point, we predict the critical budget required for a discontinuous transition for $\epsilon = 0.1$ to be $b_c^\mathrm{est} \approx 0.058$. Indeed, this is confirmed by the numerical results shown in Fig.~\ref{fig:fig3_critical_budget_parameters}: the transition is continuous for $b \le 0.05$, while the transition for $b \ge 0.06$ is already discontinuous. 

But how can the transition become discontinuous for $b > b_c$? Stopping the $\epsilon$-fraction most extreme events prevents any cluster above a certain size $C_\mathrm{thresh}$ to appear in the network. As more links are added, this threshold slowly increases. This is similar to the dynamics of the Bohman-Frieze-Wormald (BFW) model \cite{bohman04_bfw_percolation}. In fact, we observe comparable behavior in the sub-critical regime: there is a hierarchy of thresholds $p_k > 0$, $k=3,4,\hdots$ where a new largest cluster of size $S_1 = k$ first appears. 
As in the BFW model, these $p_k$ converge to constant, finite values $0 < p_k < p_c$ for large systems and announce the critical transition as $p_k \rightarrow p_c$ for $k \rightarrow \infty$ (see Supplementary Information). 
Thus, the same mechanism that leads to a discontinuous transition in the BFW model causes a discontinuous transition for optimal resource-limited control of percolation \cite{bohman04_bfw_percolation, chen13_superciritcal_explosive_bfw, chen14_microtransition}.

We have studied other control strategies and cost functions, for example cost proportional to the size of the clusters involved in the link, $c\left[S(i), S(j)\right] = S(i) + S(j)$ (see Supplementary Information).
We find for all of the studied cost functions that a small budget leads to a continuous transition, whereas a larger budget further delays the transition and eventually leads to a discontinuous transition. However, when the cost scales with the size of the clusters, the transition only becomes discontinuous when the budget scales superlinearly $B \sim \mathcal{O}(N^a)$ with $a>1$.\\

\begin{figure}
\centering
\includegraphics[width=0.4\textwidth]{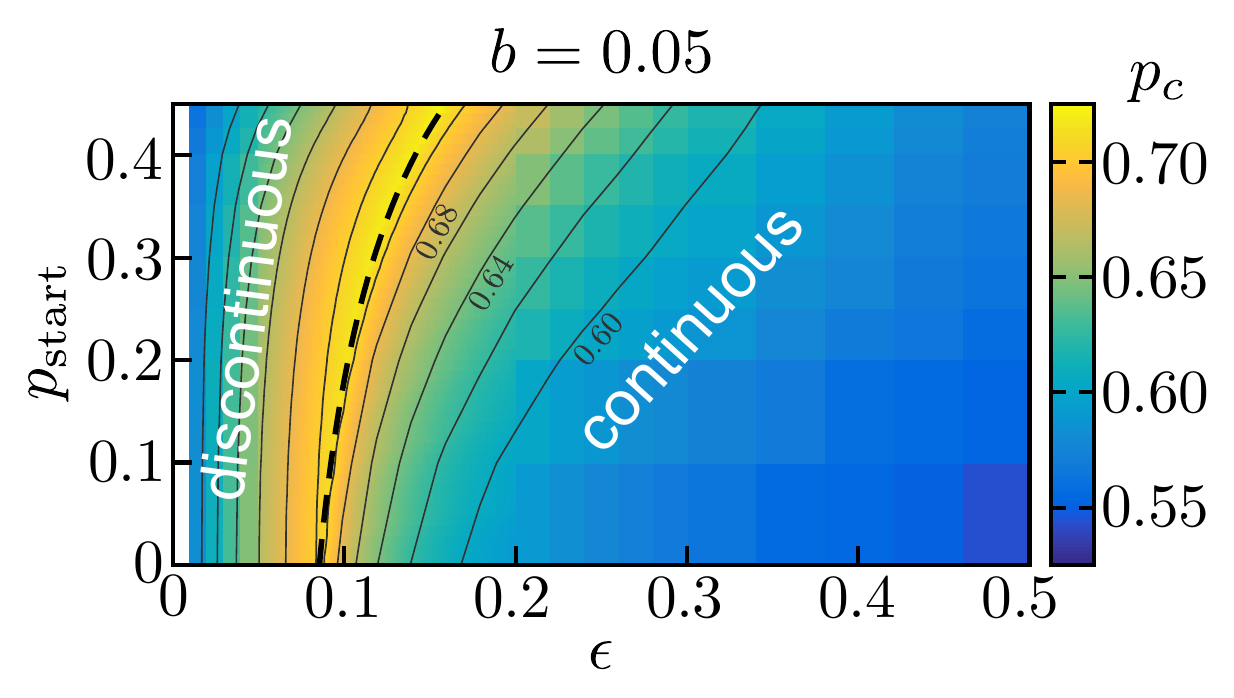}
\caption{\textbf{`Phase diagram' and discontinuous transition as a result of optimal control.}
Position of the transition $p_c$ for budget parameter $b=0.05$ as a function of intervention intensity $\epsilon$ and intervention start $p_\mathrm{start}$. Results for system size $N=2^{20}$, averaged over $R = 128$ realizations. The largest delay with $p_c \approx 0.72$ is achieved for a set of optimal intervention parameters (bright yellow) that separate the continuous from the discontinuous transition regime. The transition becomes discontinuous as a result of the optimal resource-limited control. The black dashed line represents our estimate for this optimal parameter set in ($\epsilon$, $p_\mathrm{start}$) space (see text). The thin lines indicate lines of constant $p_c$.}
\label{fig:fig4_optimal_intervention_parameters}
\end{figure}

\noindent\textbf{Limited observability.}
One realistic limitation to the control of connectivity is observability. In particular, we might not be aware of problems, such as emerging large clusters, early on in the process and only begin interventions after some time $p_\mathrm{start}$.
Under these conditions, how do we best utilize a limited budget? Adapting equation (\ref{plast}) to include $p_\mathrm{start}$ leads to the relation $b = \int_{p_\mathrm{start}}^{p_c^\mathrm{max}} \frac{\epsilon_\mathrm{eff}(p)}{1 - \epsilon_\mathrm{eff}(p)} \mathrm{d}p$ describing the optimal intervention parameters. Calculating the optimal start and intensity of the interventions with $b=0.05$ and the observed $p_c^\mathrm{max} = 0.72$, we obtain a good agreement with the numerical results in Fig.~\ref{fig:fig4_optimal_intervention_parameters}. As explained above, the line of optimal control parameters separates the regimes of continuous and discontinuous transitions. As required by the constraint of limited resources our control scheme is much more efficient than explosive percolation models at controlling percolation: We achieve $p_c = 0.72$ with only about one intervention per 15 added links, much less than comparable competitive percolation models, which reject one link for each link added (see Supplementary Information).

Interestingly, we find that for fixed intervention cost interventions close to the percolation threshold are slightly more effective than early interventions ($p_c$ slowly increases as a function of $p_\mathrm{start}$ along the critical line). This result, however, is specific to constant intervention costs as other cost functions can lead to a different behavior: interventions as early as possible, $p_\mathrm{start}=0$, are optimal for intervention costs that grow with the size of the connected clusters (see Supplementary Information).\\

\section*{Discussion}
We have derived a control strategy to efficiently delay percolation with limited resources.

In contrast to previous models constructed to delay the percolation transition \cite{achlioptas09_explosive, costa10_explosive, riordan11_explosive, nagler11_single_links, grassberger11_explosive, dsouza15_review}, we find that the transition remains smooth and in the same universality class as random percolation for non-optimal control when the resources are exhausted before the transition.
Given a fixed budget, maximal delay of the percolation transition is achieved by optimizing the control protocol such that the budget runs out exactly at the percolation threshold. 
While the percolation transition can be delayed by control interventions, this resource-optimal delay inevitably results in a discontinuous percolation transition that becomes effectively \emph{un}controllable, 
since the addition of a single link induces a macroscopic change in the connectivity.

It is commonly believed that interventions taken as early as possible can have the biggest impact to avoid large-scale connectivity \cite{helbing13_networked_risk}.
We have shown that this is not always the case: a strong effort to intervene right at the beginning can diminish the budget to such an extent that more timely interventions become impossible in crucial stages.

The framework we developed on the basis of random network growth highlights the unintended consequences of trying to control the percolation transition by delaying it \cite{helbing13_networked_risk, lee17_universal, sornette14_perpetualMoney}. Likely, similar effects will occur for other control schemes as well. This work may thus help to design control schemes in other networks, specific to the underlying network dynamics and its constraints, in particular when resources are scarce. \\

\textbf{Acknowledgments.}
We gratefully acknowledge support from the G\"ottingen Graduate School for Neurosciences and Molecular Biosciences (DFG Grant GSC 226/2 [MS]), 
the Portuguese Foundation for Science and Technology (FCT) 
under Contracts nos. UID/FIS/00618/2013, and IF/00255/2013 [NA], and 
the ETH Risk Center (RC SP 08-15) [JN].\\ 

\bibliographystyle{apsrev4-1}
\bibliography{manuscript_OptimalPercolation}

\clearpage

\onecolumngrid

{\centering \Large  \textbf{Supplementary Information}\\
accompanying the manuscript\\
\textbf{Controlling percolation with limited resources\\}}
{\center \normalsize{Malte Schr\"oder, Nuno Araujo, Didier Sornette, Jan Nagler}\flushright}
\quad\\
\quad\\

In the main manuscript we discussed the control of percolation with limited resources for interventions. In particular, we considered a growth model where, starting from an empty network with $N$ nodes and no links, at each step we choose a link $e_{ij}$ uniformly at random to add to the network. Control is implemented by a choice to prevent this link, paying a cost $c\left[S(i),S(j)\right]$ from a limited budget $B$. To decide whether we intervene in the link addition process we consider the $\epsilon$ fraction of links that, when added to the network, would create the largest clusters and prevent those links. Interventions are only possible as long as we have sufficient budget $B \ge c\left[S(i),S(j)\right]$, starting from $B = bN$. In the main manuscript we discussed the main features of this model for constant cost $c[S(i),S(j)] = 1$, showing that even a relatively small number of interventions can significantly delay the percolation transition. For small budgets this transition is continuous, but becomes discontinuous at some critical values $b_c$ and $\epsilon_c$. We also showed that the transition is maximally delayed for a given budget when the intervention intensity $\epsilon$ is exactly equal to this critical value.\\

In the following we first illustrate the necessary efficiency of our intervention rule, showing a comparison to the standard model of explosive percolation. We then present an overview and finite size scaling analysis for various parameters with $c\left[S(i),S(j)\right] = 1$, showing that the critical exponents observed for the continuous transition agree with those expected for random percolation. Thus, our control scheme does not change the universality class of the (continuous) percolation transition.

We then consider in more detail the distribution of the interventions and give a detailed derivation of the approximation describing the critical intervention parameters. We illustrate the similarity of the sub-critical behavior to the BFW (Bohman-Frieze-Wormald) model and provide further evidence that the transition indeed becomes discontinuous for large budget. Specifically, we consider unlimited budget in line with the (implicit) assumption in many other percolation rules.

Furthermore, we show examples of the reverse intervention rule, enhancing the percolation transition, with similar results.

Finally, we illustrate the robustness of our results with respect to different cost functions, among others $c\left[S(i),S(j)\right] = S(i) + S(j)$. These results are qualitatively similar and we can establish a direct mapping between results for the different cost functions. A given budget ``fixes'' the number of interventions up to small fluctuations and there is a direct correspondence between the budget and intervention parameters for both cost functions. Contrary to the results from the main manuscript, however, early interventions are more effective in this case.\\

\begin{figure}[!h]
\centering
\vspace{1cm}
\includegraphics[width=0.5\textwidth]{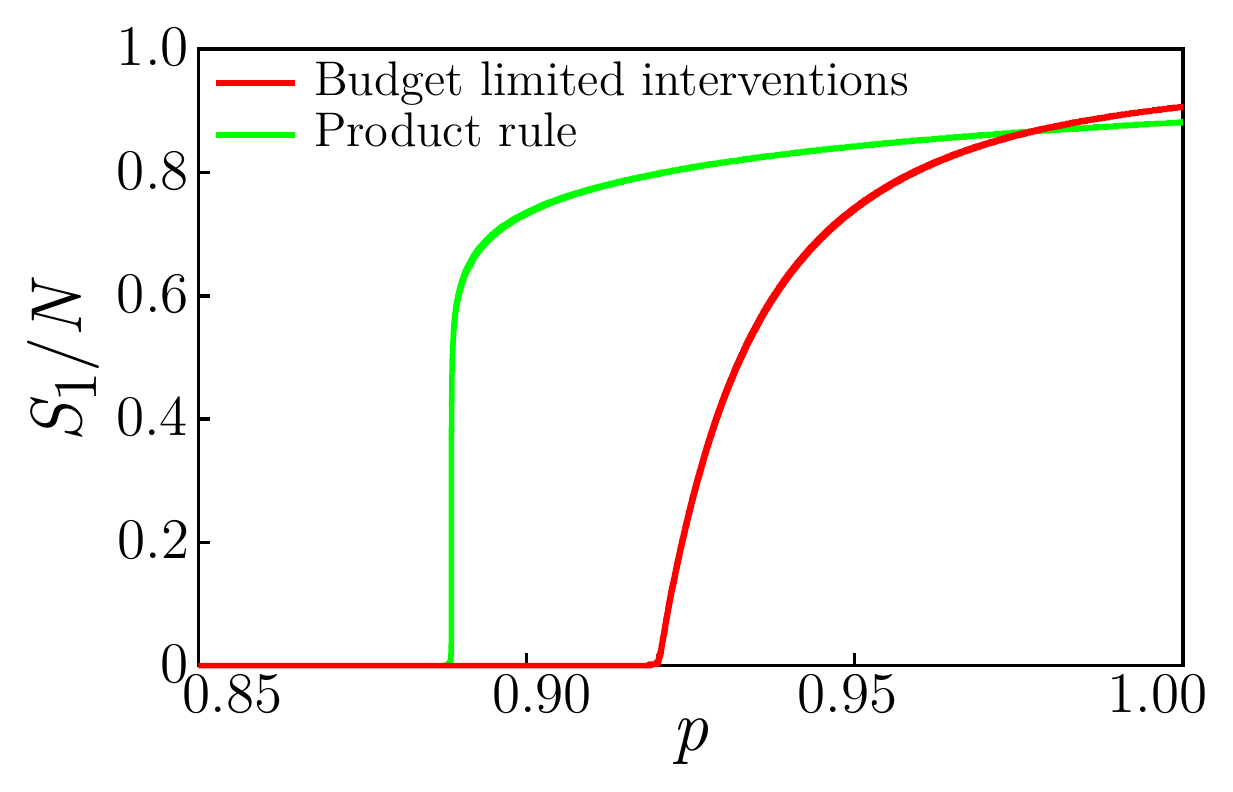}
\caption{Single realizations of the largest cluster size for the resource-limited percolation introduced in the main manuscript and the product rule \cite{achlioptas09_explosive} resulting in explosive percolation ($N=2^{25}$).  The parameters are $b = 0.88$ and $\epsilon = 0.62$. In both models $L_\mathrm{rej} \approx 0.88N$ links are rejected until the phase transition occurs. This illustrates that the intervention rule defined in the main manuscript is more effective in delaying the transition with the same budget (number of rejected links). The transition even remains continuous and does not become explosive. After the phase transition a large cluster appears quickly in both cases. However, we note that links are still being rejected in the case of the product rule slowing the growth of the macroscopic cluster in this case.}
\label{fig:explosive_percolation_comparison}
\end{figure}

\newpage
We compare the product rule of explosive percolation \cite{achlioptas09_explosive} to our intervention rule. The product rule chooses one of two links in each step, rejecting the link $e_{ij}$ with the larger product $S(i) S(j)$. This choice significantly delays the percolation transition, but results in a very abrupt, ``explosive'' transition with behavior almost indistinguishable from a discontinuous transition even in very large systems. In order to compare the product rule to our model, we first consider the ``budget'' required for the product rule: in each step one link is rejected, thus for constant cost $c[S(i),S(j)] = 1$ the product rule requires a budget $B(p) = pN$ up until $p$. Therefore, until the phase transition at $p_c^\mathrm{PR} \approx 0.889$ we need a budget a little over $B = 0.88N$. We use the same budget for our intervention rule and choose a good (though not optimal) intervention intensity $\epsilon = 0.62$. As shown in Fig.~\ref{fig:explosive_percolation_comparison} our intervention rule delays the percolation transition more efficiently while also keeping the transition continuous as random percolation (shown in detail in the next section).

\newpage
\section*{Finite size analysis}

\begin{figure}[!h]
\centering
\includegraphics[width=0.85\textwidth]{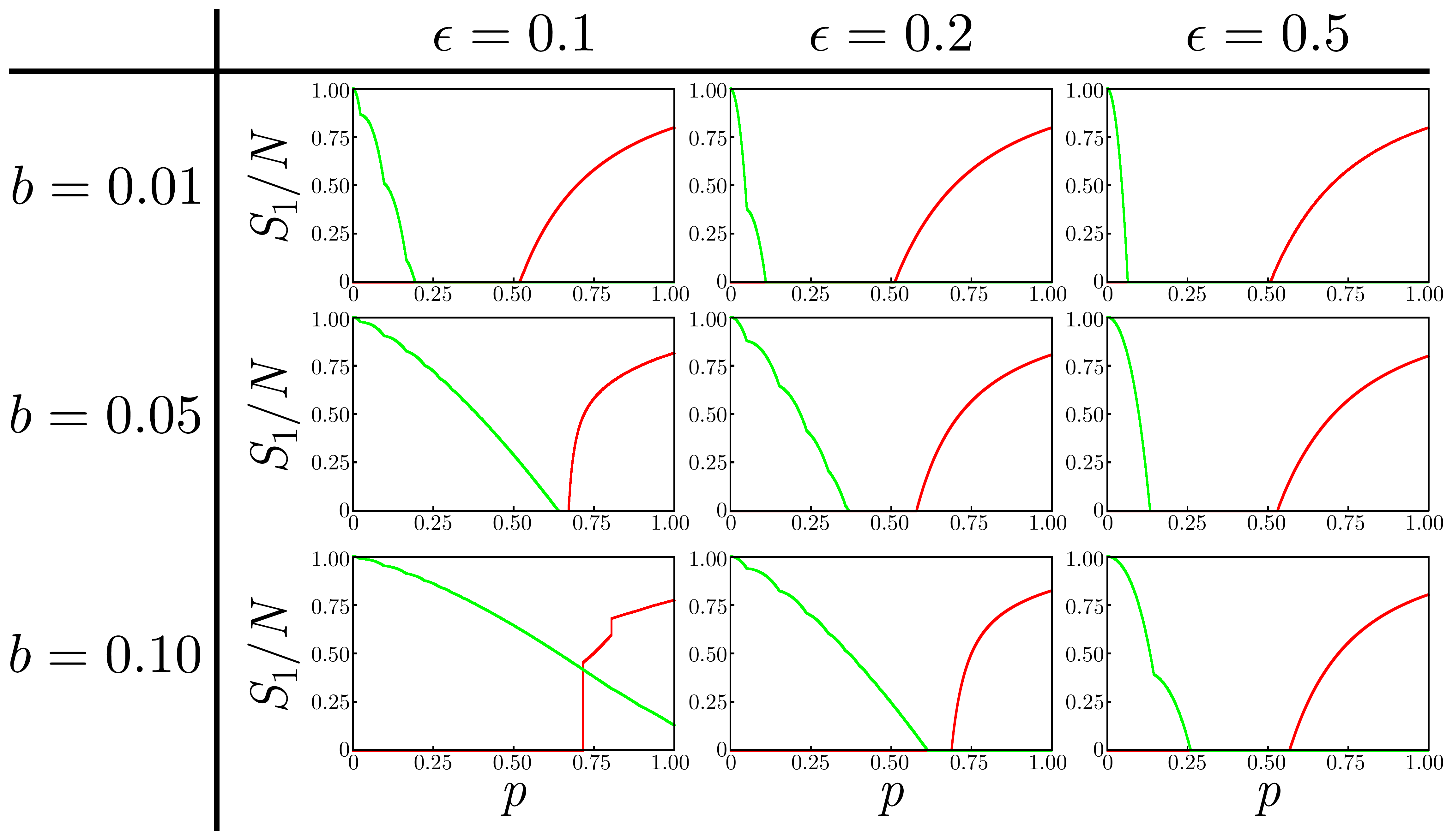}
\caption{Single realizations of the largest cluster size and the budget for various parameter combinations with $c\left[S(i),S(j)\right] = 1$ ($N=2^{25}$). The panels show the relative size of the largest cluster (red lines) and the remaining \emph{fraction} of the total budget (green line) for different initial values $B = bN$ and intervention intensities $\epsilon$. Depending on the parameters the delay between the last interventions (budget reaching $0$) and the percolation transition changes. The transition is smoothest when this gap is large. When the budget lasts until after the percolation transition, the transition is discontinuous. The kinks in the budget curve are signatures of changes in the largest cluster present in the system, see Fig.~\ref{fig:intervention_distribution} for more details.}
\label{fig:limited_budget_single_realizations_fixed_cost}
\end{figure}

Here we present the results for the finite size analysis of resource-limited percolation as discussed in the main manuscript for different parameters. Fig.~\ref{fig:limited_budget_single_realizations_fixed_cost} shows single realizations for parameter combinations used in the finite size scaling analysis. The transition is continuous as long as the interventions end before the transition, as described in the main manuscript, unless the budget is large and the intervention intensity is small ($b = 0.1, \epsilon = 0.1$).

\begin{figure}[!h]
\centering
\includegraphics[width=0.9\textwidth]{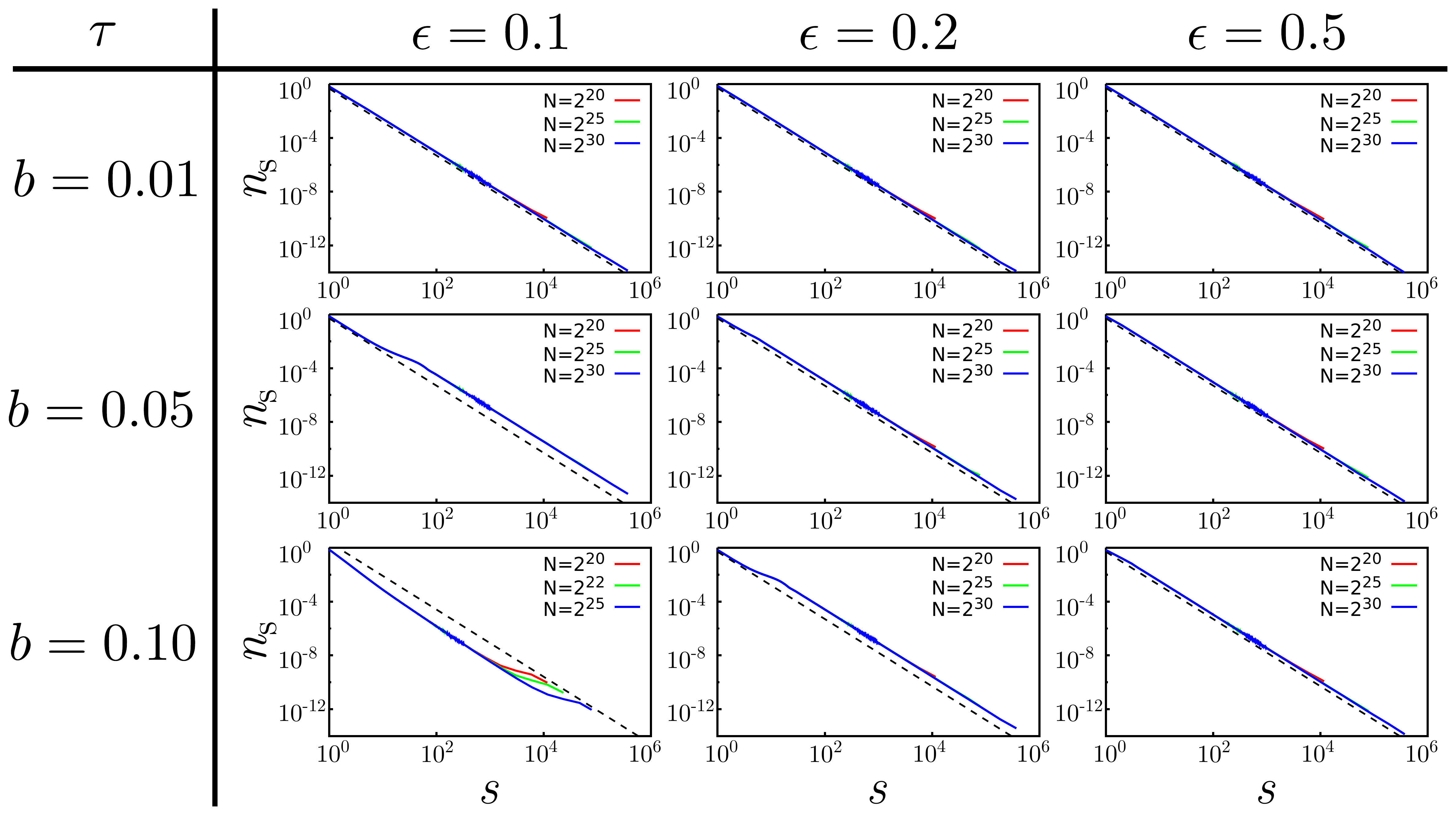}
\caption{Cluster size distribution at $p_c$ for different parameter values with $c\left[S(i),S(j)\right] = 1$ and various system sizes $N$ ($N=2^{20}$, $2^{25}$ and $2^{30}$ averaged over $1024$, $256$ and $64$ realizations, respectively [$N=2^{20}$, $2^{22}$, and $2^{25}$ for $b=0.1$, $\epsilon=0.1$]). Here, $n_S$ describes the \emph{relative} frequency of clusters of size $S$. The scaling is expected to follow a power law $n_S \sim S^{-\tau}$ for large $S$. The dashed black lines show the scaling expected for random percolation with exponent $\tau = 5/2$ (not normalized). The peak in the cluster size distribution for small $S$ is a signature of the finite size of clusters in the system when the interventions stop. Larger budgets allow for more interventions shifting the peak to larger $S$ and making it more pronounced. Higher intervention intensities use the budget earlier, shifting the peak to lower $S$.}
\label{fig:csd_scaling}
\end{figure}

\begin{figure}[!h]
\centering
\includegraphics[width=0.9\textwidth]{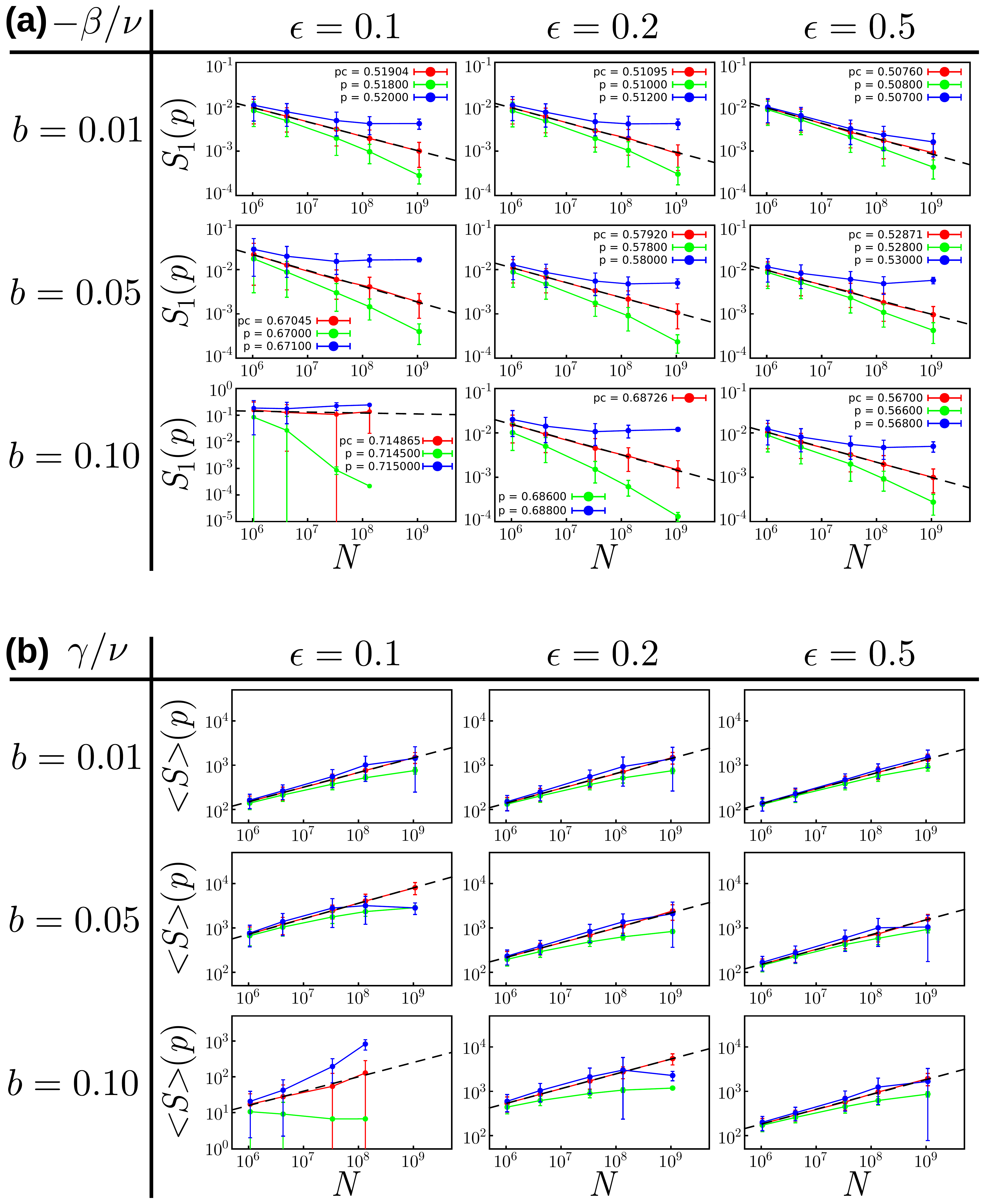}
\caption{Results of the finite size scaling analysis for $c\left[S(i),S(j)\right] = 1$ assuming standard critical scaling  $S_1(p) \sim \left|p-p_c\right|^\beta$ and $\left<S\right>(p) \sim \left|p-p_c\right|^\gamma$. The error bars indicate the standard deviation, averages are over $1024$ to $64$ realizations. (a) Results of the finite size scaling analysis for the exponent $-\beta/\nu \approx -1/3$ (detailed results in Tab.~\ref{tab:finite_size}), showing the same behavior as expected for random percolation for all continuous transitions. For $b=0.1$, $\epsilon = 0.1$ we find $\beta \approx 0$, corresponding to a discontinuous transition. (b) Results of the finite size scaling analysis for the exponent $\gamma/\nu \approx 1/3$ (detailed results in Tab.~\ref{tab:finite_size}), showing the same behavior as expected for random percolation for all continuous transitions. 
The individual panels show the data at the (expected) critical point in red and two other values slightly above and below for comparison (blue and green, respectively). Lines are guides to the eye. The black dashed line shows the best fit.
}
\label{fig:finite_size_scaling}
\end{figure}

We performed a finite size scaling analysis, assuming the scaling $S_1(p) \sim \left|p-p_c\right|^\beta$ and $\left<S\right>(p) \sim \left|p-p_c\right|^{-\gamma}$ for the size of the largest cluster and the mean cluster size, respectively, as well as the correlation length $\xi(p) \sim \left|p-p_c\right|^{-\nu}$. The resulting exponents are listed in Tab.~\ref{tab:finite_size} and corresponding figures are shown in Fig.~\ref{fig:finite_size_scaling}. For all continuous transitions the exponents agree well with those expected for random percolation. Additionally, the cluster size distribution at the critical point, shown in Fig.~\ref{fig:csd_scaling}, decays with the same exponent $\tau = 5/2$ as for random percolation, evidencing that controlled percolation remains in the same universality class as long as the transition is continuous. In particular, the transition never becomes ``explosive'' or ``weakly discontinuous''.

We note that, while the results for $\epsilon = 0.1$ and $b=0.1$ agree with a discontinuous transition, they are likely less accurate than those for the continuous case. For the specific parameter combination we observe two transition very close by that only resolve for very large system sizes (see Fig.~\ref{fig:unlimited_budget_transition_average} below), making it difficult to accurately determine the critical point and measure only the first transition.

\begin{table}[!h]
\centering
\begin{tabular}{|c|c|c|c|}
\multicolumn{4}{l}{$-\beta/\nu$} \\
\hline
 $b$\textbackslash$\epsilon$ & $0.1$ & $0.2$ & $0.5$	\\
\hline
 $0.01$ & $-0.325(3)$ & $-0.338(8)$ & $-0.337(9)$\\
\hline
 $0.05$ & $-0.35(1)$ & $-0.336(3)$ & $-0.333(8)$\\
\hline
 $0.10$ & $-0.03(5)$ & $-0.338(6)$ & $-0.337(5)$\\
 \hline
\end{tabular}
\hspace{0.1\textwidth}
\begin{tabular}{|c|c|c|c|}
\multicolumn{4}{l}{$\gamma/\nu$} \\
\hline
 $b$\textbackslash$\epsilon$ & $0.1$ & $0.2$ & $0.5$	\\
\hline
 $0.01$ & $0.331(3)$ & $0.338(5)$ & $0.331(2)$\\
\hline
 $0.05$ & $0.347(5)$ & $0.343(7)$ & $0.333(7)$\\
\hline
 $0.10$ & $0.40(5)$ & $0.334(7)$ & $0.339(5)$\\
 \hline
\end{tabular}
\caption{Exponents $-\beta/\nu$ (left) and $\gamma/\nu$ (right) found by finite size scaling analysis. The corresponding figures are shown in Fig.~\ref{fig:finite_size_scaling}. The values agree with the exponents expected for random percolation $-\beta/\nu = -1/3$ and $\gamma/\nu = 1/3$ when the interventions end before the transition. For $\epsilon = 0.1$ and $b=0.1$ the result is consistent with the expected $\beta = 0$ of a discontinuous transition.}
\label{tab:finite_size}
\end{table}

\clearpage
\section*{Predicting $p_\mathrm{last}$ and optimal intervention parameters}

\begin{figure}[!h]
\centering
\includegraphics[width=0.66\textwidth]{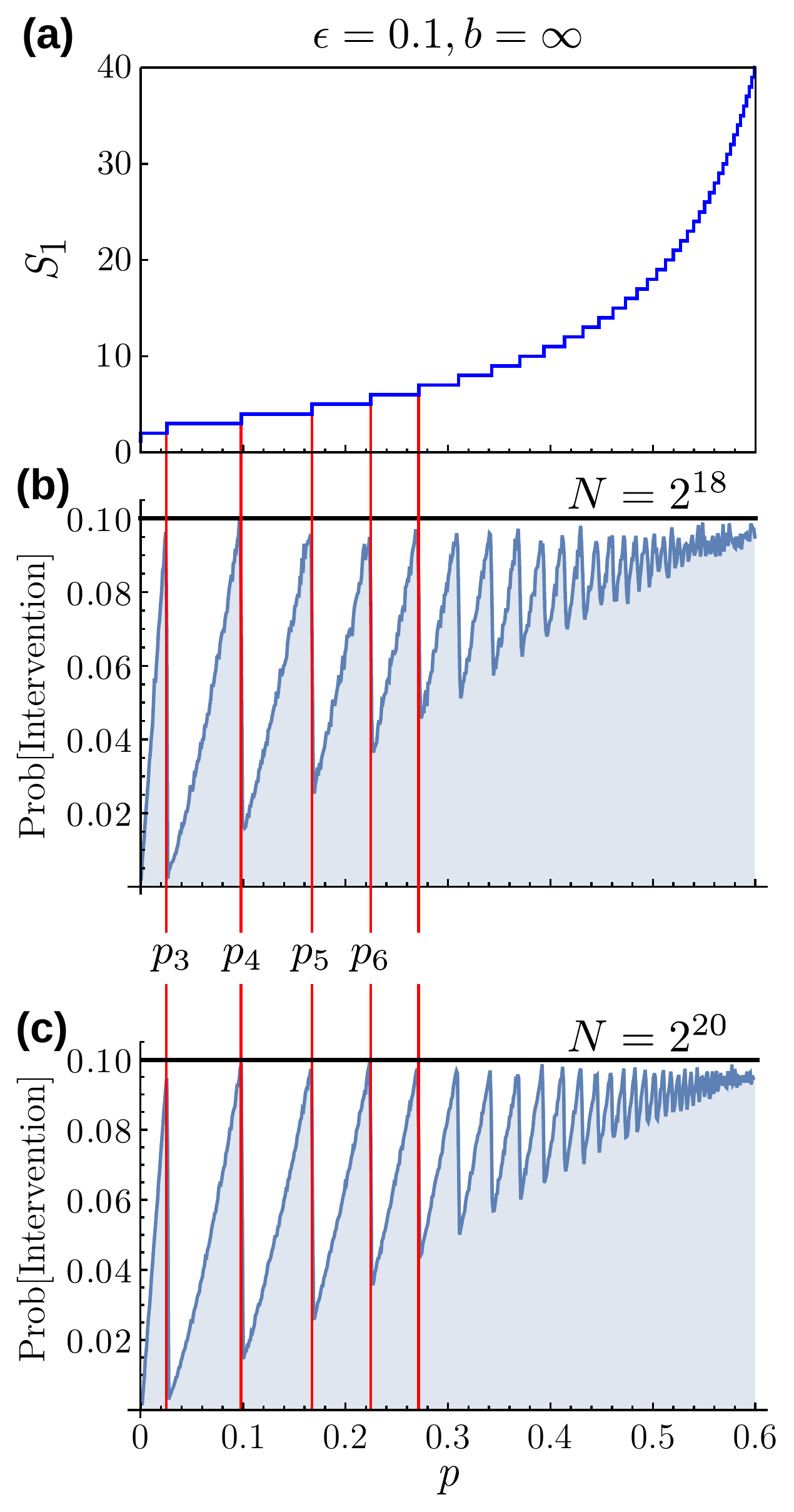}
\caption{(a) Single realization of the largest cluster size in the sub-critical regime for $N=2^{25}$ with unlimited budget and $\epsilon = 0.1$. (b,c) Probability of an intervention for a single link chosen at $p$ for two different system sizes averaged over $100$ realizations each. Initially, intervention probability is low, since a cluster of size at least $2$ is almost always created, but the chance to create a cluster of size $3$ or larger (which would trigger an intervention) is much smaller than $\epsilon$. This probability slowly increases with the addition of more links. When a new cluster size appears in the network, creating clusters larger than this size is again less likely than $\epsilon$ and the intervention probability ``resets''. This causes the transitions to $S_1 = 3$ at $p_3$ (and so on) to occur at fixed positions [compare also to the single realization in panel (a)]. This behavior is very similar to the sub-critical evolution of the largest cluster in the BFW (Bohman-Frieze-Wormald) model, leading to a discontinuous transition at $p_k \rightarrow p_c$ for $k \rightarrow \infty$.}
\label{fig:intervention_distribution}
\end{figure}

\begin{figure}[!h]
\centering
\includegraphics[width=0.5\textwidth]{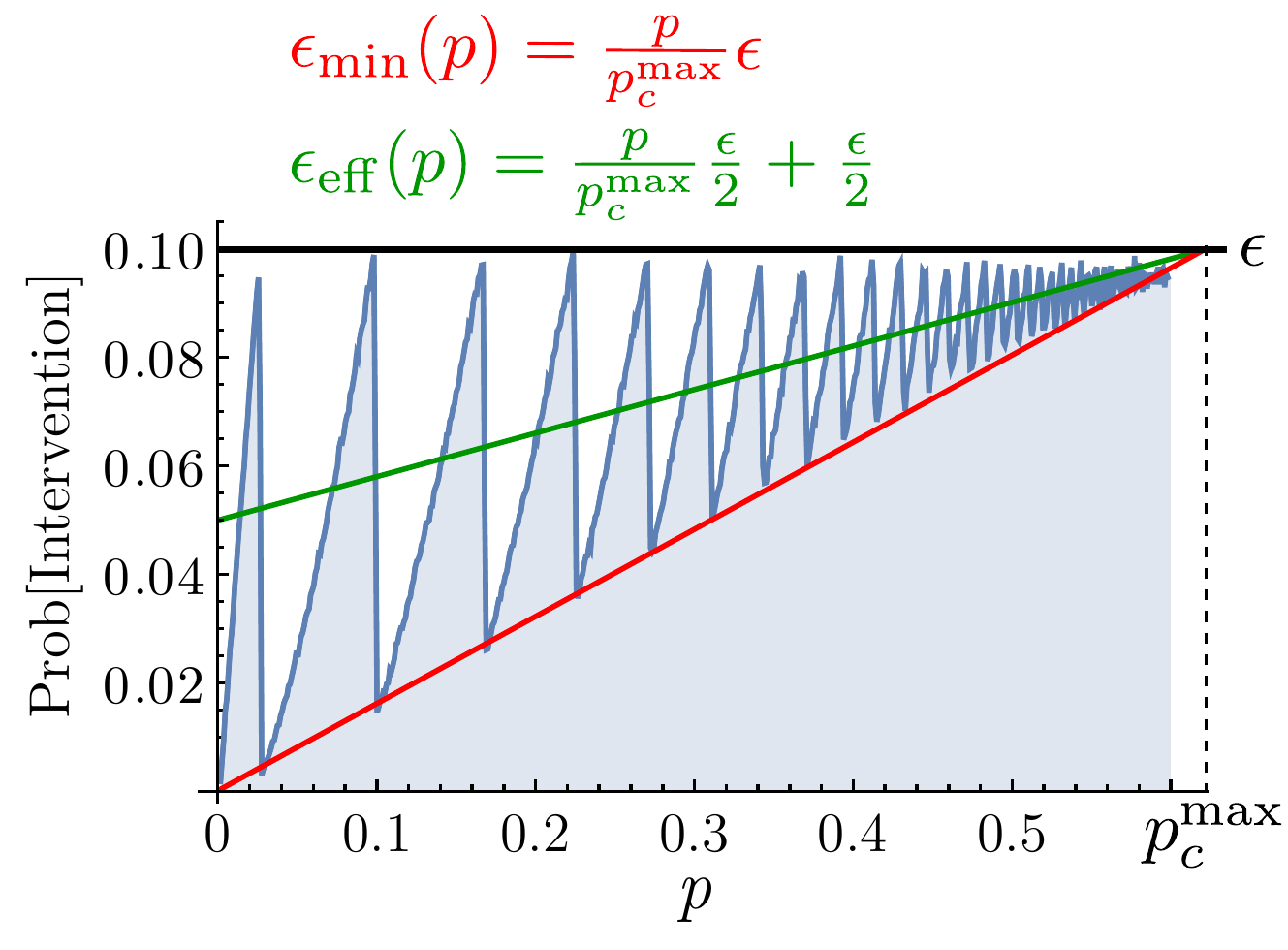}
\caption{Probability of an intervention for a single link chosen at $p$ for system size $N=2^{20}$ averaged over $100$ realizations. The red and green lines illustrate the motivation for introducing the effective intervention rate $\epsilon_\mathrm{eff}$ (green), describing a local average of the true intervention rate (here for $p_\mathrm{start} = 0$).}
\label{fig:effective_intervention_rate}
\end{figure}

In the main manuscript we discussed the estimation of $p_\mathrm{last}$ to predict when the interventions will end. This empirical approximation allowed us to accurately predict the critical intervention parameters, where the transition becomes discontinuous. Here, we present a detailed derivation of this approximation.

To begin, we first note that the intervention probability is not constant and equal to $\epsilon$, as one might have expected from the definition of the intervention rule. As mentioned already in the main manuscript, the interventions are distributed non-uniformly. This is easiest to understand by considering the first link: we never prevent the first link since the probability to create a cluster of size $2$ is $\mathrm{Prob}\left[S(k) + S(l) \ge 2\right] = 1 > \epsilon$. Thus the probability of an intervention $\epsilon(p = 0) = 0$. Similarly, the first few links are unlikely to be prevented, since a link creating a cluster of size $3$ or larger is chosen with vanishing probability.

In fact, one can think about the intervention rule in the following way: We always prevent the most extreme links. This is equivalent to preventing all clusters above a certain size (until these links become too likely). This means, when the size of the largest cluster just changed to $S_1$, the probability to create a larger cluster is usually smaller than $\epsilon$. However, the links creating a cluster of size $S_1$ are not prevented as the probability to create a cluster larger \emph{or equal} to $S_1$ is larger than $\epsilon$. Thus, after these micro-transitions of the largest cluster size, the intervention probability drops. In fact, we find that these transitions to a new largest cluster size happen at well defined times, constant across different system sizes (see Fig.~\ref{fig:intervention_distribution}). This behavior is very similar to the sub-critical evolution of the BFW (Bohman-Frieze-Wormald) model \cite{bohman04_bfw_percolation,chen14_microtransition}. This similarity also supports the discontinuity of the transition as in the BFW model, when the interventions last until after the critical point.\\

We can use this observation to derive an (empirical) estimate for the budget used for interventions up to $p$. To reiterate the basic idea described in the main manuscript, given a constant intervention rate $\epsilon$ we would expect $\Delta L_\mathrm{int} = \epsilon \Delta L_\mathrm{total}$ interventions to occur during the sampling of $\Delta L_\mathrm{total}$ links. During this period, we add only $N \Delta p = \Delta L = (1-\epsilon) \Delta L_\mathrm{total}$ links. With constant intervention costs, the number of interventions directly correspond to the budget used during this interval and we obtain $\Delta b = \Delta B / N = \frac{\epsilon}{1 - \epsilon} \Delta p$.

However, as discussed above, the true intervention rate is not constant. For an accurate estimation we need to use a varying intervention rate. Since we do not know the exact form of $\epsilon(p)$, we use an empirically determined ``effective intervention rate'' $\epsilon_\mathrm{eff}(p)$, describing a local average of $\epsilon(p)$ (illustrated in Fig.~\ref{fig:effective_intervention_rate} for $p_\mathrm{start} = 0$). This intervention rate depends on the intervention parameter $\epsilon$ and the position $p_c^\mathrm{max}$ of the critical point of the process with unlimited budget. In principle we could use the same function when $p_\mathrm{start} > 0$, however, due to the uncontrolled evolution, the intervention rate will likely be larger. To accommodate for this, we assume that the effective intervention rate at $p_\mathrm{start}$ is $\epsilon_\mathrm{eff}(p_\mathrm{start}) = p_\mathrm{start} \epsilon + \epsilon/2$ (the value obtained with $p_c^\mathrm{max} = 1/2$ for the equation in Fig.~\ref{fig:effective_intervention_rate}, as percolation is uncontrolled before $p_\mathrm{start}$). Directly at and after $p_c$ the effective intervention rate is $\epsilon_\mathrm{eff}(p \ge p_c) = \epsilon$. Together this gives
\begin{equation}\label{eq:eff_rate}
	\epsilon_\mathrm{eff}(p) = \begin{cases}	0	&\quad\mathrm{for}\quad p < p_\mathrm{start} \\
												\frac{p - p_\mathrm{start}}{p_c^\mathrm{max} - p_\mathrm{start}} \left( \epsilon - \epsilon/2 - p_\mathrm{start}\epsilon\right) + \epsilon/2 + p_\mathrm{start}\epsilon &\quad\mathrm{for}\quad p_\mathrm{start} \le p < p_c^\mathrm{max}\\
												\epsilon &\quad\mathrm{for}\quad p_c^\mathrm{max} \le p
								\end{cases}
\end{equation}

We can now use the argument we gave above and integrate Eq.~(\ref{eq:eff_rate}) over all interventions to find the total budget used. We arrive at the approximate relation
\begin{eqnarray}\label{plast}
b &=& \int_{p_\mathrm{start}}^{p_\mathrm{last}} \frac{\epsilon_\mathrm{eff}(p)}{1 - \epsilon_\mathrm{eff}(p)} \mathrm{d}p \\
  &=& p_\mathrm{start} - p_\mathrm{last} - \frac{ 2(p_\mathrm{start} - p_c^\mathrm{max}) \mathrm{Log}\left( \frac{2 (p_c^\mathrm{max} - p_\mathrm{start}) + \epsilon ( p_\mathrm{last} + p_c^\mathrm{max}-2p_\mathrm{start}(1+p_\mathrm{last}-p_c^\mathrm{max}) ) }{(p_c^\mathrm{max} - p_\mathrm{start})(\epsilon + 2 p_\mathrm{start} \epsilon - 2)} \right) }{ (2 p_\mathrm{start} - 1)\epsilon },
\end{eqnarray}
assuming again $p_c^\mathrm{max} \ge p_\mathrm{last}$ is the critical point of the process with parameters $\epsilon$ and $p_\mathrm{start}$ given unlimited budget.\\

Substituting $p_\mathrm{last} = p_c^\mathrm{max}$ gives the condition for optimal intervention parameters, which can be solved numerically (see Fig.~4 in the main manuscript). The quality of the estimate deteriorates for very large values of $\epsilon$ and $p_\mathrm{start}$ and small values of $b$, where interventions occur only in a small interval and averaging to $\epsilon_\mathrm{eff}$ becomes inaccurate. For the same reason, the effective intervention rate is a good approximation when estimating the optimal intervention parameters, where interventions last until $p_c$ and the error from averaging is small.

\clearpage
\section*{Unlimited budget --- discontinuous transition}

\begin{figure}[!h]
\centering
\includegraphics[width=0.66\textwidth]{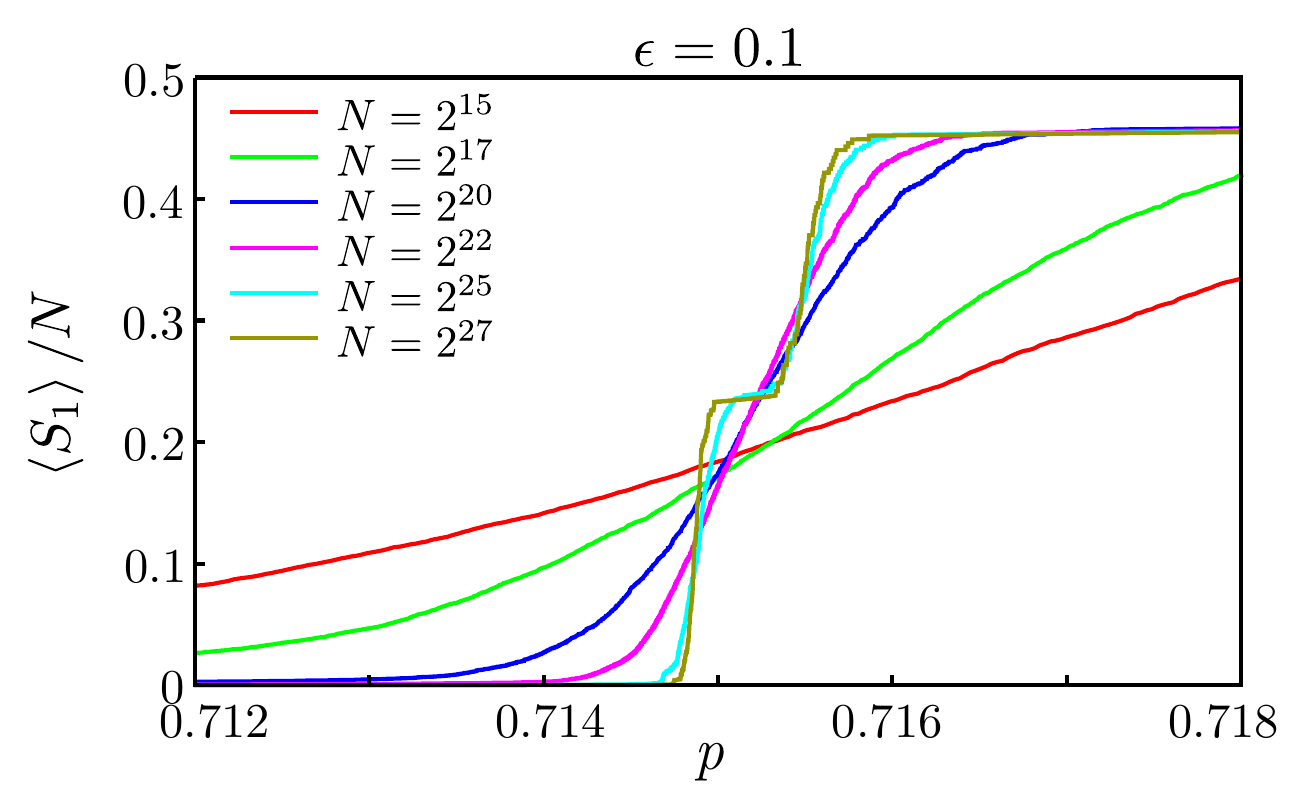}
\caption{Average size of the largest cluster during the transition for interventions with unlimited budget and $\epsilon = 0.1$. Averages are taken over $1024$ to $64$ realizations. This figure illustrates the difficulty in studying properties of the transition across different realizations: the transition is blurred out even for large finite systems, and the double transition is only revealed for very large systems.}
\label{fig:unlimited_budget_transition_average}
\end{figure}

\begin{figure}[!h]
\centering
\includegraphics[width=0.66\textwidth]{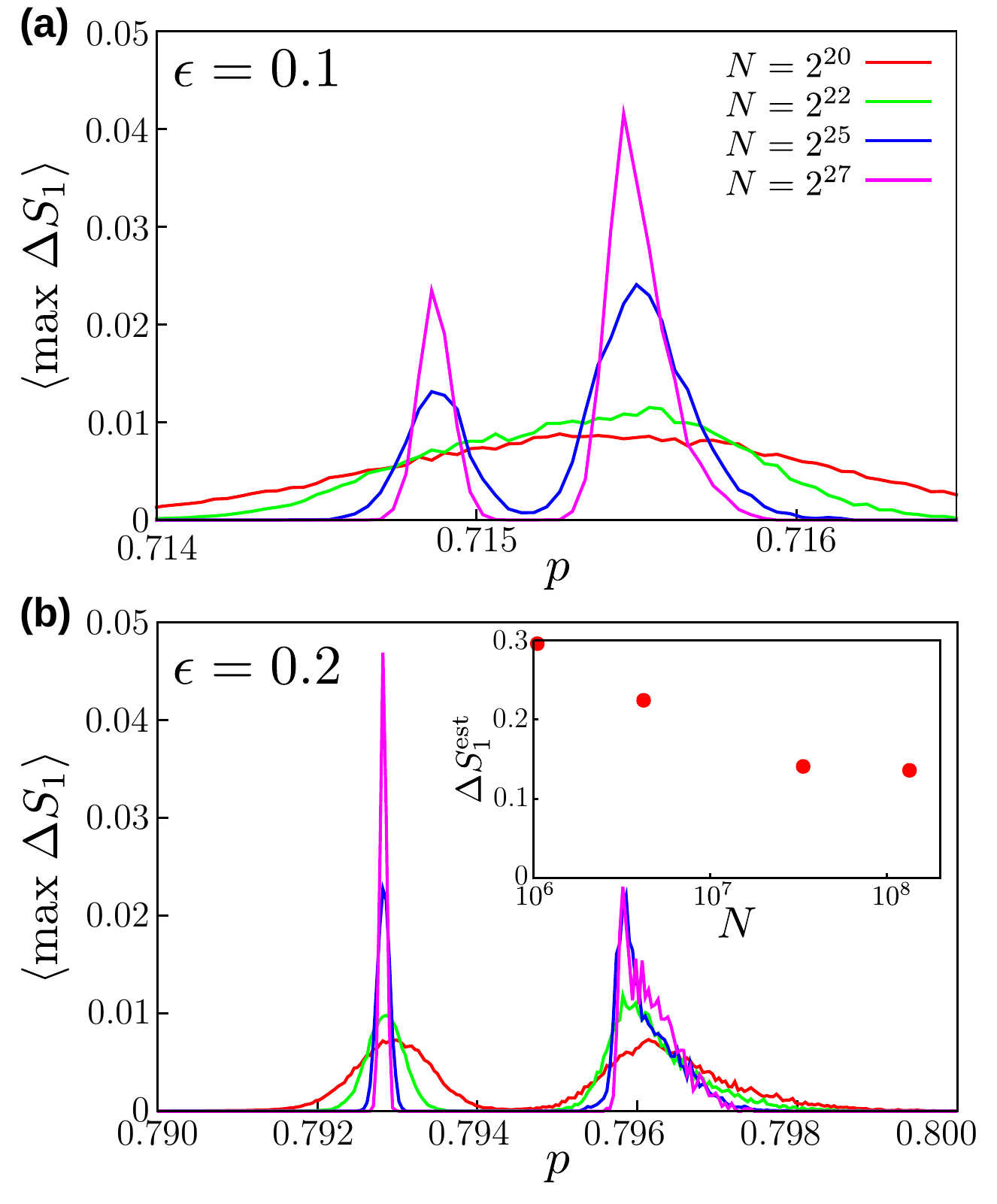}
\caption{Average maximum gap in the size of the largest cluster over $p$. The figures show the maximum size of the largest cluster in intervals $\Delta =  4 \cdot 10^{-5}$, averaged over $16384$ to $1024$ realizations for $\epsilon = 0.1$ [panel (a)] and $\epsilon = 0.2$ [panel (b)]. Results show that even though the jumps are initially indistinguishable, only two distinct jumps appear for large systems. An estimation of the expected size $\Delta S_1$ of the first jump for $\epsilon = 0.2$ shows that it is becoming constant for large systems (inset, see text for more details), further evidencing that the transition is discontinuous.}
\label{fig:unlimited_budget_max_gap}
\end{figure}

To clearly show the discontinuity of the transition when the budget survives until the transition we now consider interventions with an unlimited budget in more detail. We study the largest gap of the largest cluster, following a method from \cite{schrenk12_bfw_lattice} to resolve multiple jumps of the size of the largest cluster. We divide the region around the transition into intervals of width $\Delta p = 4 \cdot 10^{-5}$ and record the largest jump in each of these intervals (Fig.~\ref{fig:unlimited_budget_max_gap}). While the transition is blurred out for small systems, a double transition is revealed for larger system sizes. The same behavior can also be seen for averages over the multiple realizations shown in Fig.~\ref{fig:unlimited_budget_transition_average}.

We find that the largest gap of the first transition does not decay for increasing system size (even taking into account the smaller spread and thus expected larger averages). In the case of $\epsilon = 0.2$, we can quantify this by assuming a single large jump of size $\Delta S_1$ and negligible contributions of all other changes for a given realization (this approximation becomes better the larger the system). If this assumption is correct, the resulting average should simply be the product of the size of the jump $\Delta S_1$ and the probability that the jump occurs in a given interval. Consequently, we can (approximately) determine the size of the jump $\Delta S_1$ by fitting a Gaussian distribution multiplied by $\Delta S_1$ to the measured average jump size. We find $\Delta S_1^\mathrm{est} \approx 0.3$ for the smallest system $N=2^{20}$ decaying to only $\Delta S_1^\mathrm{est} \approx 0.12$ for the larger systems $N=2^{25}$ and $2^{27}$. The fact that $\Delta S_1^\mathrm{est}$ does not decay to zero shows that the jump is indeed macroscopic and the transition is discontinuous (Fig.~\ref{fig:unlimited_budget_max_gap}, inset).\\

\clearpage
\section*{Enhancing percolation}
So far, we have discussed only interventions to delay the percolation transition. Interestingly, the same logic describing the optimal intervention strategy also applies to enhance percolation. Instead of stopping the $\epsilon$-fraction most extreme events, we simply stop the $\epsilon$-fraction least extreme events (specifically including links connecting nodes in the same cluster). We use constant intervention costs $c[S(i),S(j)] = 1$, as in the main manuscript. Again, optimal interventions for constant costs necessarily end at the percolation threshold. Interventions lasting longer have no additional effect on the threshold and interventions ending earlier create an extensive interval of uncontrolled percolation before the transition, partially negating the effect of the interventions.

\begin{figure}[!h]
\centering
\includegraphics[width=\textwidth]{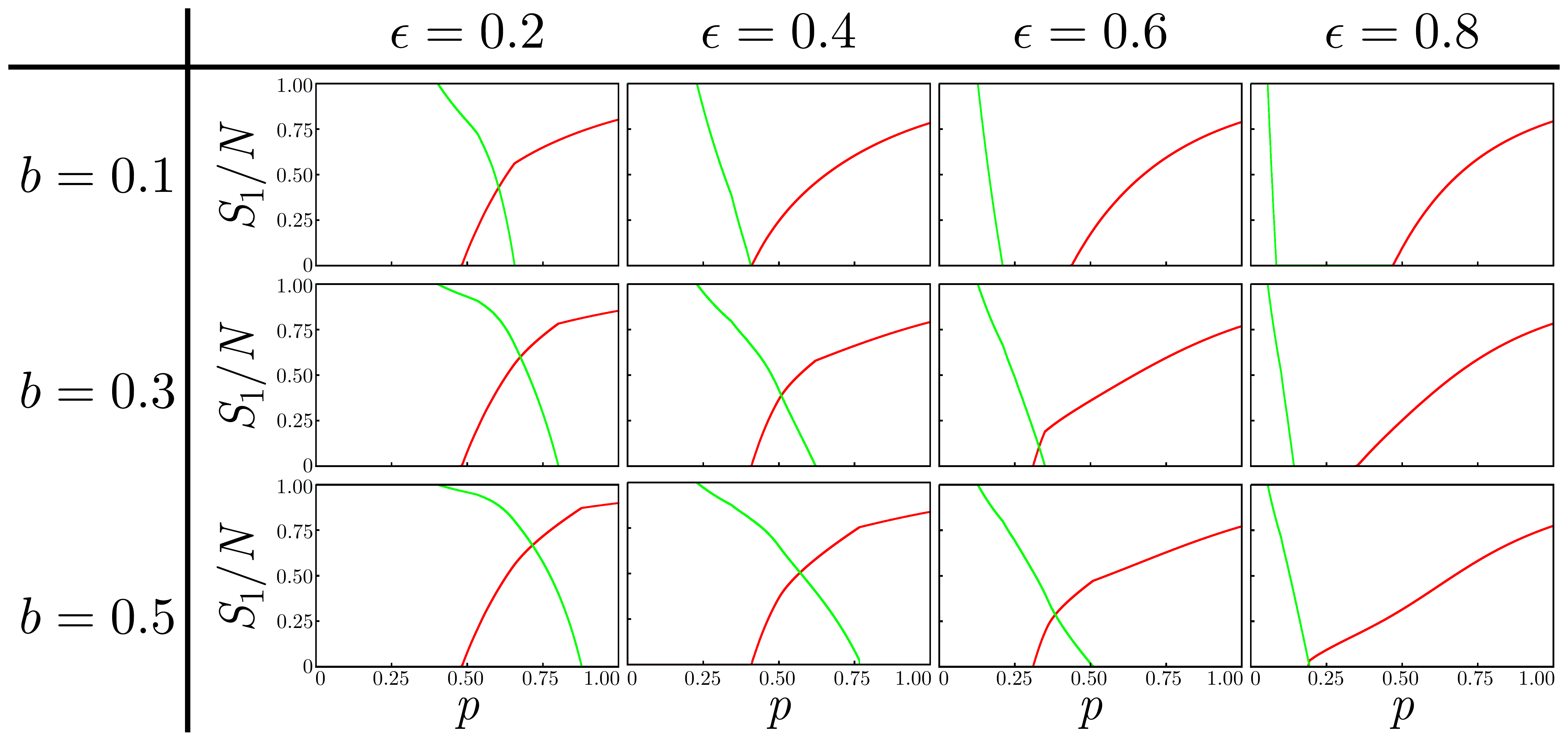}
\caption{Single realizations of the largest cluster size and the budget for various parameter combinations and $c\left[S(i),S(j)\right] = 1$ ($N=2^{25}$) for interventions enhancing percolation. The panels show the relative size of the largest cluster (red lines) and the remaining \emph{fraction} of the total budget (green line) for different initial values $B = bN$ and intervention intensities $\epsilon$ (stopping the $\epsilon$-fraction least extreme events). Depending on the parameters the transition is enhanced more or less strongly. As for delaying the transition, interventions are most efficient, when the interventions last exactly until the transition.}
\label{fig:limited_budget_single_realizations_enhancing}
\end{figure}

In Fig.~\ref{fig:limited_budget_single_realizations_enhancing} we show examples for single realizations of percolation enhancing interventions. The results confirm that the effect is largest ($p_c$ is smallest when the budget runs out exactly at $p_\mathrm{last} = p_c$. Interestingly, the transition also becomes steeper at $p_c$ when the interventions last until (after) the percolation threshold. Thus, also in the case of enhancing percolation, optimal control inherently leads to reduced controllability of the transition.

\clearpage
\section*{Resource-limited control of percolation under various cost functions}

In order to illustrate the universality of our results, in the following we consider control of percolation using different cost functions as well as intervention rules based on other observables. First, we discuss a theoretical argument, why our results naturally extend to different cost functions.

Due to the self-averaging behavior of the percolation model, the number of interventions for a given cost function $c\left[S(i),S(j)\right]$ with given parameters $\epsilon$ and $b$ is fixed with a negligible variance (relative to the system size). For large systems we thus find a direct correspondence to a system with constant cost $c'=1$ and parameters $\epsilon'$ and $b' = b'(b, \epsilon)$, where $b'N$ is simply given as the average number of interventions.\\

\begin{figure}[!h]
\centering
\includegraphics[width=0.85\textwidth]{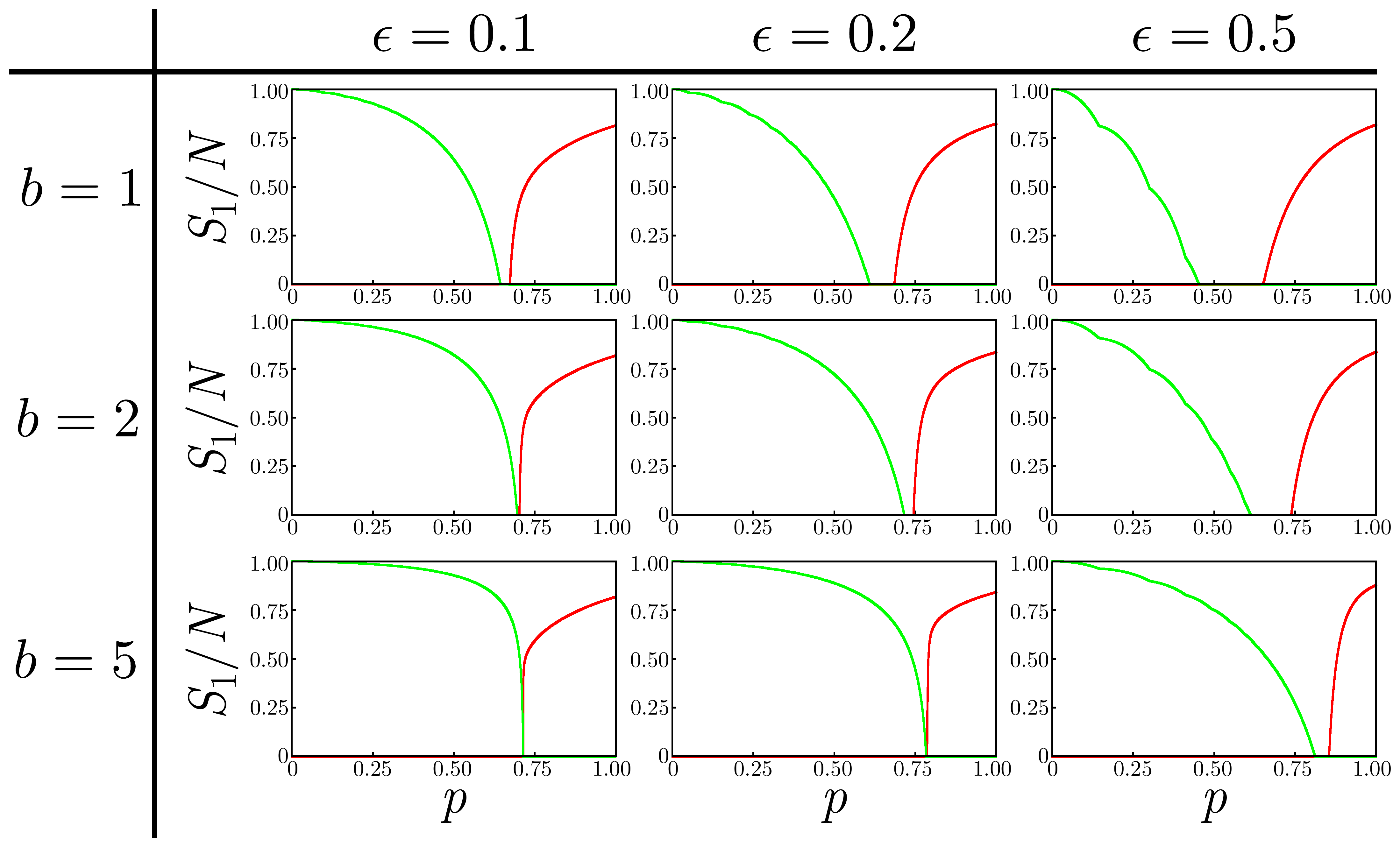}
\caption{Single realizations of the largest cluster size and the budget for various parameter combinations with $c\left[S(i),S(j)\right] = S(i) + S(j)$ for system size $N=2^{25}$. The panels show the relative size of the largest cluster (red lines) and the remaining \emph{fraction} of the total budget (green line) for different initial values $B = bN$ and intervention intensities $\epsilon$. Depending on the parameters the delay between the last interventions (budget reaching $0$) and the percolation transition changes. The transition is smoothest when this gap is large. As discussed in the text, the transition is always continuous, since the budget runs out before the transition.}
\label{fig:limited_budget_single_realizations}
\end{figure}

We mostly discuss results using the cost function $c\left[S(i),S(j)\right] = S(i) + S(j)$, where the cost of an intervention scales with the size of the clusters involved. In this case, for any budget $B = bN$ with constant $b$ we always observe a continuous transition (this means the corresponding $b' < b'_c$). This is illustrated in an overview of single realizations in Fig.~\ref{fig:limited_budget_single_realizations}. This can be understood with the following (rough) argument: consider the average budget used for interventions up to a point $p$
\begin{eqnarray}
\left<b(p)\right> &=& 1/N \left< \sum_\mathrm{Interventions} S(i) + S(j) \right> \nonumber\\
&\le& 1/N \sum_\mathrm{Interventions} 2\left< S \right> \lesssim 2 \epsilon \int_0^p <S>(p') \mathrm{d}p' \,.
\end{eqnarray}
With the standard assumption for the critical scaling $<S>(p') \sim \left| p_c-p' \right|^{-\gamma}$ this integral is finite for all $p < p_c$ but diverges at $p_c$. Thus any constant, finite budget $b$ will run out at some point $p_\mathrm{last} < p_c$, regardless of the value of $\epsilon$, and the transition will be continuous. Conversely, we can reach any $p_\mathrm{last} < p_c$ with a finite budget $B = \mathcal{O}(N)$. We can thus establish a direct mapping between the two cost functions with $b \in \left[0,\infty\right)$ for $c\left[S(i),S(j)\right] = S(i) + S(j)$ and $b' \in \left[0, b'_c\right)$ for constant cost (main manuscript). A delayed start of the interventions at $p_\mathrm{start}$ does not qualitatively change this mapping.

\newpage

Due to the different cost of interventions, it becomes much more important to intervene early, when interventions are cheap. Finding the optimal $\epsilon$ for a given budget now does not mean keeping the interventions up the longest: if the intensity is too large we prevent relatively unimportant links. If the intensity is too small, some interventions are executed close to the critical point and are very costly, reducing the total number of interventions. The optimal delay is obtained for intermediate values of $\epsilon$, balancing the observed effectiveness of interventions close to the critical point (see main manuscript) with the increasing costs.

\begin{figure}[!h]
\centering
\includegraphics[width=0.66\textwidth]{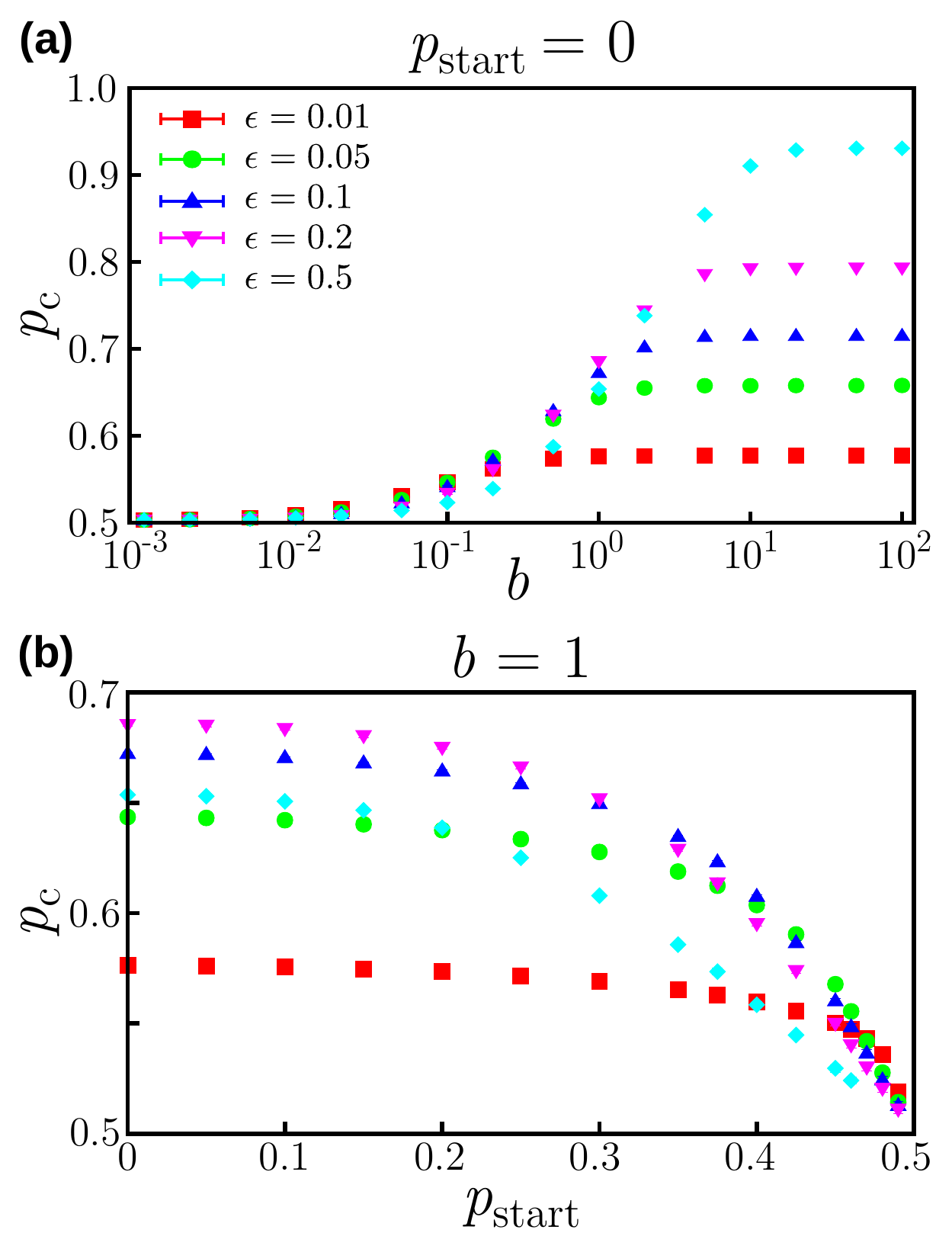}
\caption{Position of the percolation transition for various parameter combinations and cost function $c\left[S(i),S(j)\right] = S(i) + S(j)$, averaged over $256$ realizations of $N=2^{25}$. Error bars indicating the standard deviation are smaller than the symbol size. (a)~$p_c$~when interventions are possible for all $p$ ($p_\mathrm{start} = 0$). Obviously, a larger budget allows more interventions and leads to larger $p_c$. Considering a fixed budget, it is clearly visible that different values of the intervention intensity $\epsilon$ are optimal, e.g., large $\epsilon$ are feasible for large budgets, while they are sub-optimal for smaller budgets. (b) Resulting $p_c$ for the same $\epsilon$ versus $p_\mathrm{start}$, now for fixed $b=1$. Clearly, starting the interventions earlier always results in a larger $p_c$. Considering a fixed $p_\mathrm{start}$, the optimal intervention intensity $\epsilon$ changes: small intensities are sub-optimal when interventions are possible early but become optimal as $p_\mathrm{start}$ approaches $p_c^\mathrm{ER} = 1/2$.}
\label{fig:pc_vs_parameters}
\end{figure}

\begin{figure}[!h]
\centering
\includegraphics[width=0.66\textwidth]{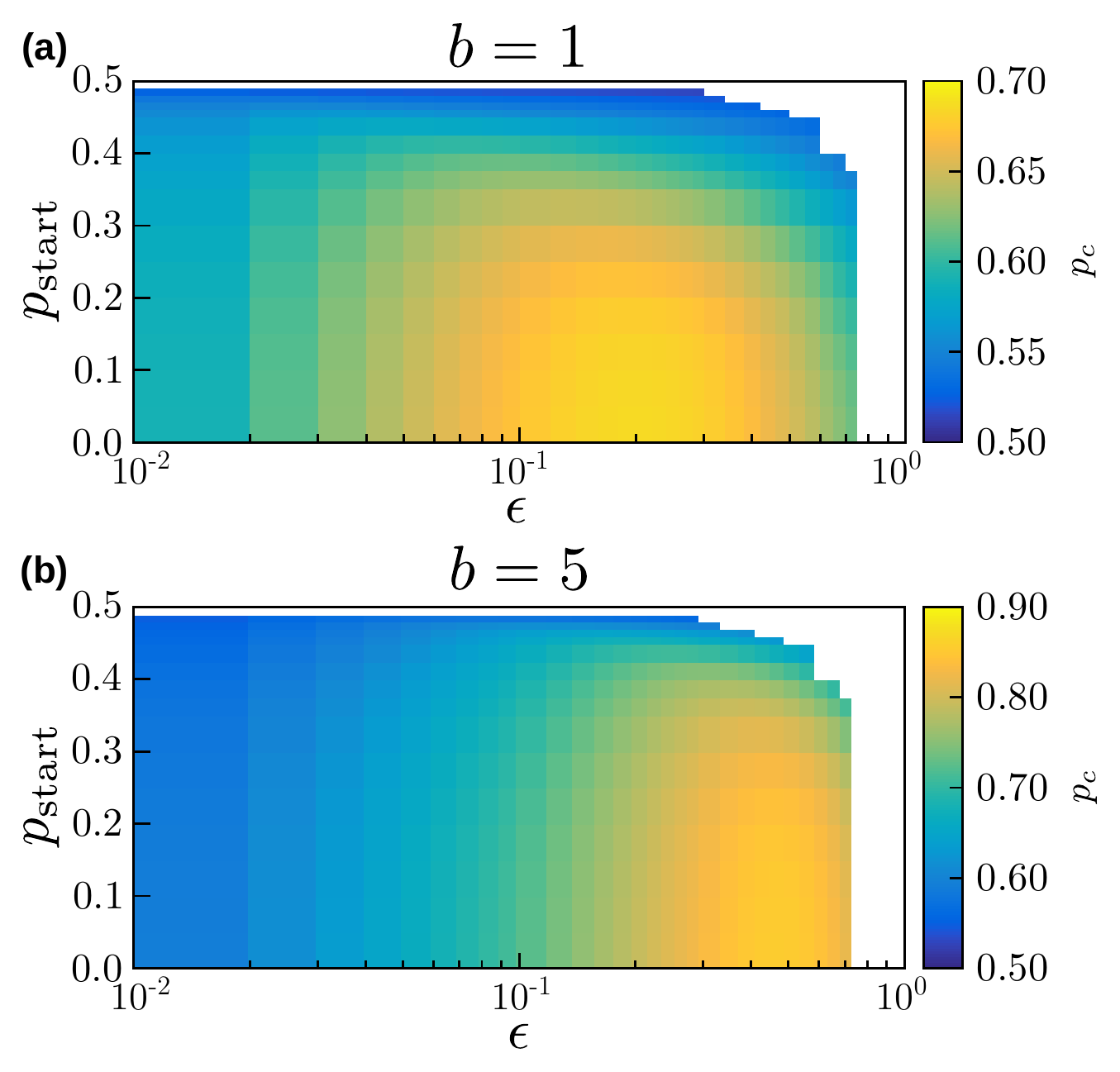}
\caption{Position of the phase transition $p_c$ for cost function $c\left[S(i),S(j)\right] = S(i) + S(j)$ and interventions with $b=1$ [panel (a)] and $b=5$ [panel (b)] and parameters $\epsilon$ and $p_\mathrm{start}$ ($N=^{25}$), averaged over $256$ realizations. The qualitative behavior is identical in both cases. Early interventions with intermediate intensity are optimal. Since more budget is available for $b=5$, the optimal intervention intensity as well as the possible delay is larger than for $b=1$. }
\label{fig:pc_vs_parameters_colormap}
\end{figure}

The resulting delay of the percolation transition for various parameters, shown in detail in Fig.~\ref{fig:pc_vs_parameters} and \ref{fig:pc_vs_parameters_colormap}, illustrates the findings summarized above: (i) a larger budget will always increase the delay of the percolation transition (Fig.~\ref{fig:pc_vs_parameters}a), (ii) starting the interventions early and (iii) using an intermediate intensity results in the largest delay of $p_c$ (Fig.~\ref{fig:pc_vs_parameters}b, \ref{fig:pc_vs_parameters_colormap}).

\begin{figure}[!h]
\centering
\includegraphics[width=0.66\textwidth]{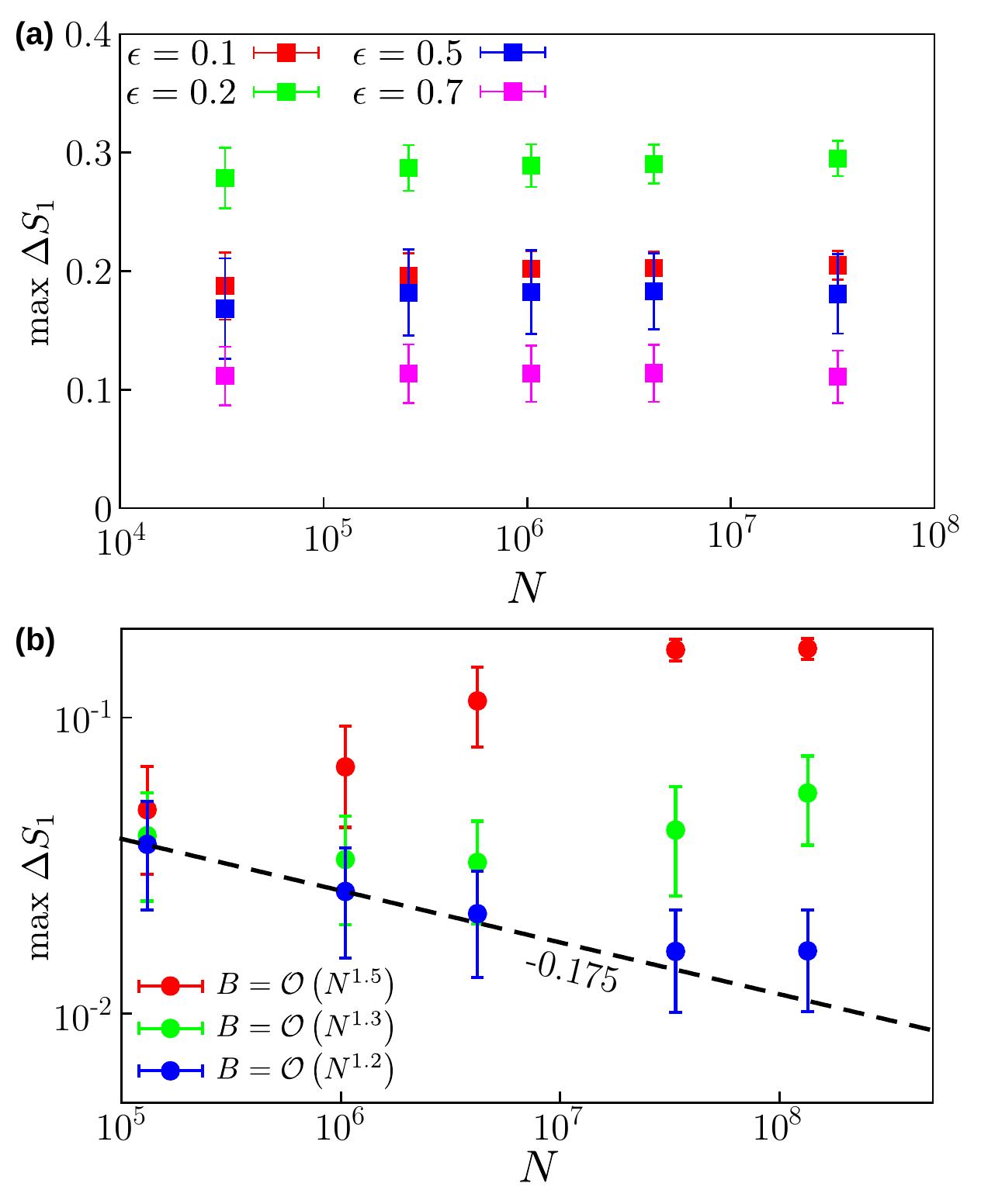}
\caption{Discontinuous transition with super-linear budget scaling for cost function $c\left[S(i),S(j)\right] = S(i) + S(j)$. Error bars indicate the standard deviation, averages are taken over $1024$ to $64$ realizations. (a) The largest gap is constant with increasing system size for unlimited budget and different values of the intervention intensity, evidencing a discontinuous transition. (b) The largest gap does not disappear with increasing system size for superlinear budget scaling. The black line shows a power law scaling expected for a continuous transition (slope chosen by eye to approximate the data for small systems). While finite size effects make it difficult to study the behavior for arbitrary super-linear scaling, it is clear that for all $B \ge \mathcal{O}\left(N^{1.2}\right)$ the largest gap in the size of the largest cluster does not disappear, evidencing a discontinuous transition.}
\label{fig:unlimited_budget_superlinear_max_gap}
\end{figure}

However, we recover the discontinuous transition observed in the main manuscript for superlinear budget scaling $B \sim \mathcal{O}(N^a)$ with $a>1$ (Fig.~\ref{fig:unlimited_budget_superlinear_max_gap}). This is required for interventions to last until (after) the percolation transition, where a single intervention will (likely) cost an extensive amount $\Delta B = \mathcal{O}(N)$. Specifically, considering the scaling of the  cluster sizes before the transition, we can expect a critical budget on the order of $B_c \sim \mathcal{O}\left[N \log(N)\right]$. However, strong finite size effects make this prediction impossible to verify numerically.

\begin{figure}[!h]
\centering
\includegraphics[width=\textwidth]{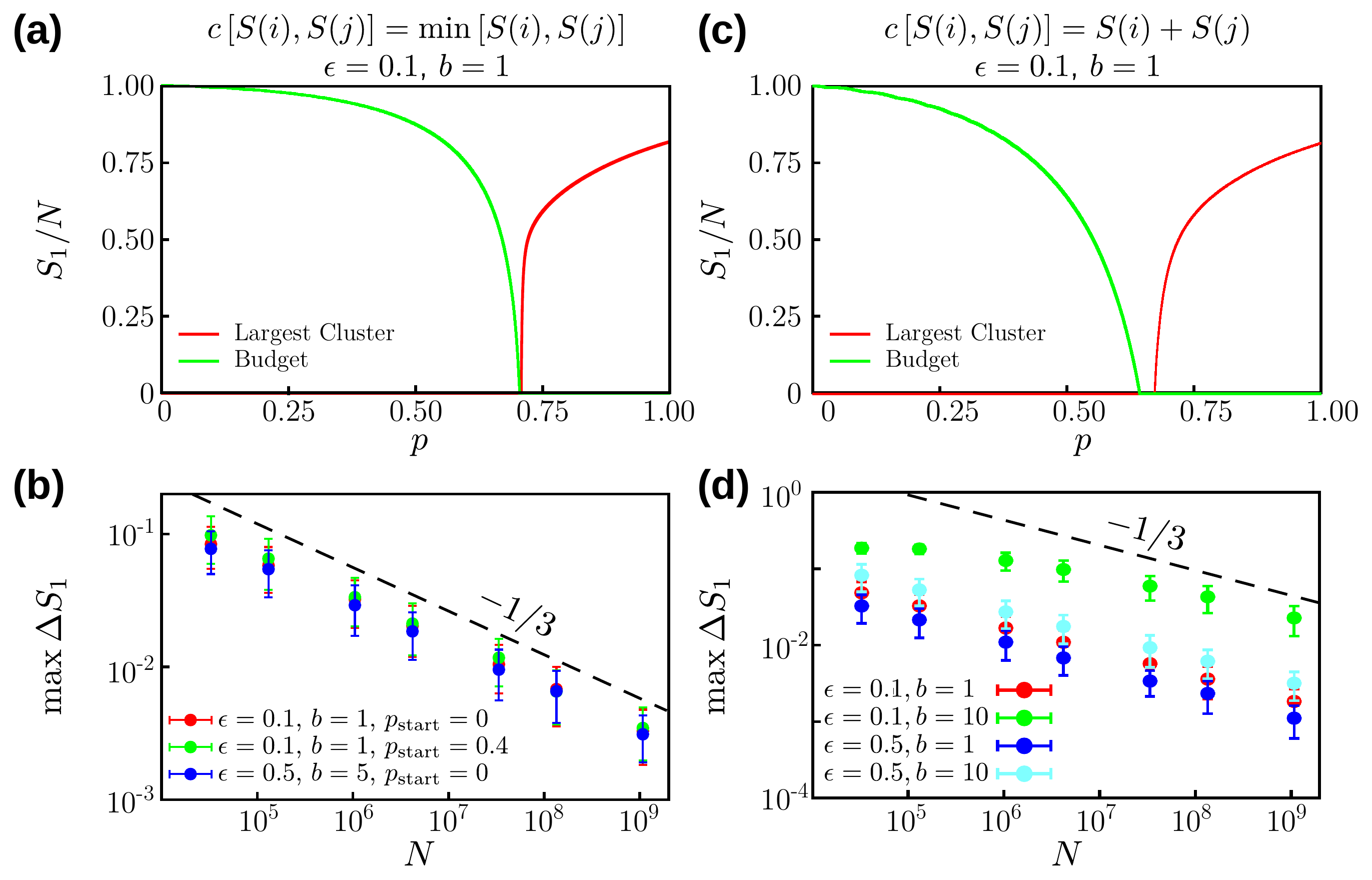}
\caption{Results for the intervention rule as used in the main manuscript, but for different intervention costs as noted above the two columns ($N=2^{25}$).  Error bars indicate the standard deviation, averages are taken over $1024$ to $64$ realizations, depending on the system size. (a,c) Single realizations of the size of the largest cluster as well as the remaining fraction of the total budget. (b,d) The largest gap of the size of the largest cluster for various parameter combinations. The transition is continuous and behaves as expected for random percolation in all cases.}
\label{fig:other_cost}
\end{figure}

\clearpage
To further illustrate the generality of these results, depending only on the scaling but not the specific choice of the cost function, we considered other intervention cost functions, specifically $c\left[S(i),S(j)\right] = \min\left[S(i), S(j)\right]$ and $c\left[S(i),S(j)\right] = S(i) + S(j)$ as above (Fig.~\ref{fig:other_cost}), as well as $c\left[S(i),S(j)\right] = \max\left[S(i), S(j)\right]$ (not shown) with equivalent results. Note that cost scale linearly with the size of the clusters in all cases. In all cases we can map the parameters to corresponding parameters for constant intervention cost with a budget $b' < b'_c$ as discussed above. \\

\newpage

\begin{figure}[!h]
\centering
\includegraphics[width=0.8\textwidth]{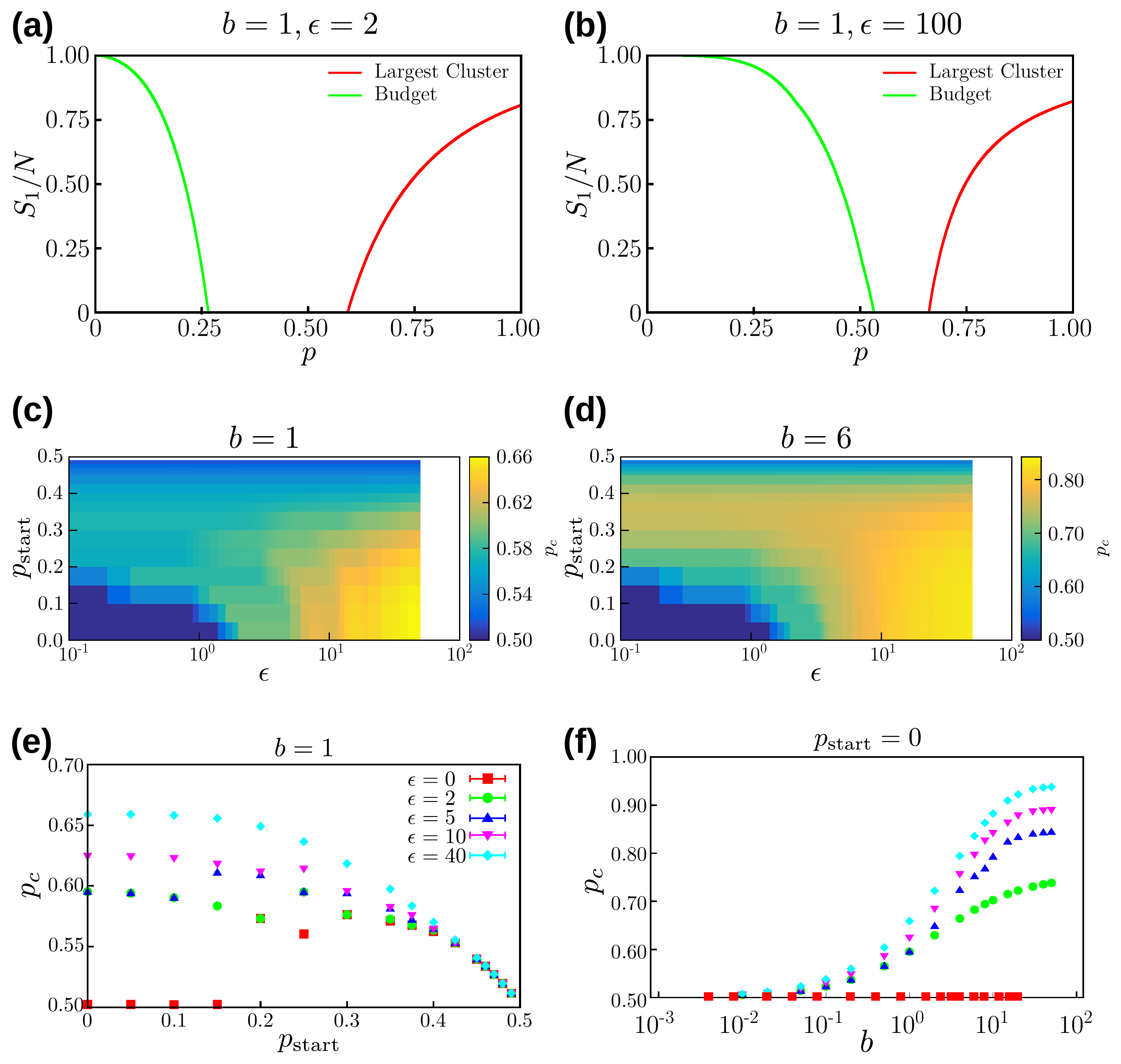}
\caption{Results for the variance intervention rule (see text) for $N=2^{25}$. Error bars indicate the standard deviation, averages are taken over $256$ realizations. (a,b) show two examples of single realizations for different intervention thresholds.\newline
(c,d) show the same colormap plots as in Fig.~\ref{fig:pc_vs_parameters_colormap}, illustrating the position of the phase transition versus different parameters: due to the cost function, early interventions are preferable.\newline
(e,f) show the resulting position of the percolation transition versus the start of the interventions and the budget, respectively (compare Fig.~\ref{fig:pc_vs_parameters}). We find the same qualitative behavior: early interventions are optimal and a larger budget obviously allows for a larger delay. Non-monotonicities in the resulting curves for $p_c$ are due to the non-monotonous scaling of the changes in the variance. The simplest example is the following: for $\epsilon = 0$ we prevent the first link when $p_\mathrm{start} = 0$ and thus use all budget at $p = 0$. However, when $p_\mathrm{start} > 0$ there are links which actually reduce the variance and our interventions are not useless.}
\label{fig:variance_rule}
\end{figure}

\begin{figure}[!h]
\centering
\includegraphics[width=0.5\textwidth]{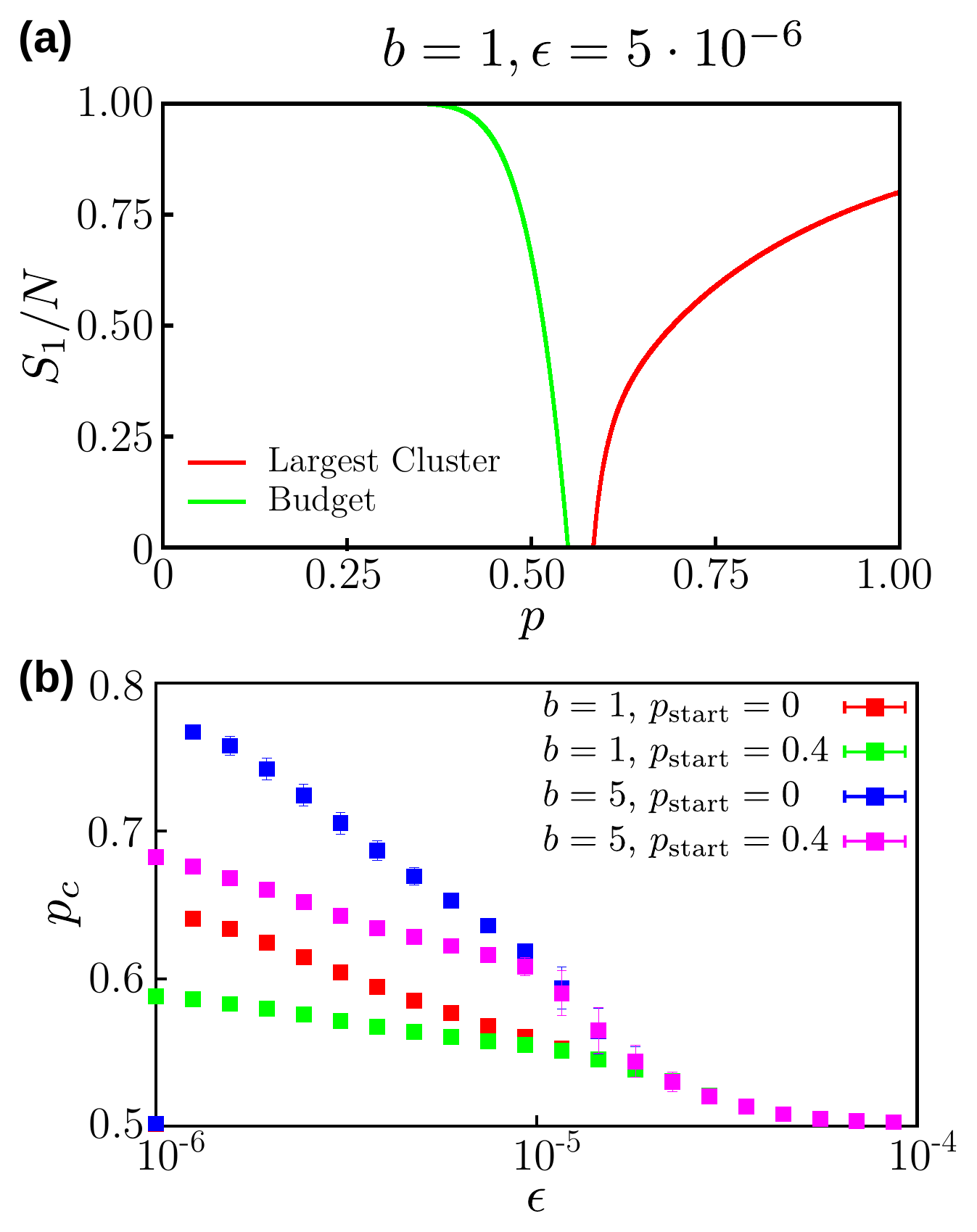}
\caption{Results for the entropy intervention rule (see text) for $N=2^{25}$. Error bars indicate the standard deviation, averages are taken over $1024$ realizations. Results are qualitatively similar to the other interventions rules considered: a larger budget will always delay the transition and, due to the cost function, early interventions are preferable. As for the variance interventions, it is possible to choose a parameter $\epsilon$ that always stops the first merger at $p=0$ (datapoints at $\epsilon = 10^{-6}$, $p_c \approx 0.5$).}
\label{fig:entropy_rule}
\end{figure}

Similarly, we consider different intervention rules. In the main manuscript we derived the intervention rule using an intuitive argument to achieve at least a given effectiveness of the intervention. Here we explicitly demonstrate other intervention rules (Fig.~\ref{fig:variance_rule} and \ref{fig:entropy_rule}), showing that the qualitative behavior is similar. We again consider intervention cost proportional to the size of the clusters $c\left[S(i),S(j)\right] = S(i) + S(j)$.

Specifically, we consider an intervention rule based on the variance of the cluster size distribution: we prevent a link if the change $\Delta V$ of the variance is larger than a certain threshold $\epsilon / N$, in order to keep cluster sizes in the network similar (and thus prevent large clusters). While this rule is less complex numerically, as we can track the variance as the network grows, it is also less efficient than the protocol derived in the main manuscript. Additionally, the threshold does not easily scale with the system size: the scaling changes depending on the shape of the cluster size distribution at any given time.

Similarly, we consider the entropy $E = \sum_S n_S log(n_S)$ instead of the variance, where $n_S$ is the probability that a random cluster has size $S$. In both cases we find qualitatively similar results as above: interventions should be applied early and with an intermediate threshold (intensity) adjusted to the budget.

\end{document}